\documentclass{article}
\usepackage{style}
\usepackage{pdfpages}
\allowdisplaybreaks

\title{ Collisional relaxation in shielded dipolar molecular gases }
\author[1]{ Reuben R. W. Wang }
\author[2]{John L. Bohn}
\affil[1]{\normalsize\textit{ITAMP, Center for Astrophysics $|$ Harvard \& Smithsonian, Cambridge, Massachusetts 02138, USA}}
\affil[2]{\normalsize\textit{JILA, NIST, and Department of Physics, University of Colorado, Boulder, Colorado 80309, USA}}
\date{}

\begin{document}

\maketitle
%
% \abstract{ 
We discuss the influence of collisions on the dynamics of an ultracold gas whose constituents interact via dipolar forces.  This dynamics is governed by the elastic scattering cross section of the molecules, which is to some extent under the experimentalist's control.  We compare side-by-side several different situations, highlighting their similarities and differences.  These situations are collisions between: 1) point dipoles; 2) electric-field-shielded polar molecules; and 3) microwave-shielded polar molecules, including the effect of microwave ellipticity.
% }

\section{ Introduction }

A fundamental issue in the preparation and study of ultracold gases is their return to equilibrium, when this equilibrium is disturbed.  For a thermal (i.e., non-quantum-degenerate) gas, equilibration is achieved due to collisions among the constituent molecules of the gas. Thus measurements of relaxation time can determine collision cross sections, in a kind of microscopic collider experiment.  Further, the approach to true quantum degeneracy is often obtained by evaporative cooling, where collisions redistribute energy to bring the gas to its lowest temperature.

Understanding the role of collisions in equilibration is therefore essential in ultracold gases.  A new frontier in this area consists of ultracold gases  of ground state molecules, whose collisions are influenced by the anisotropy of their interactions.  Part of the technology that enables these investigations is the ability to tune long-range interactions between the molecules, by means of electric or microwave fields, primarily to prevent their destruction by chemical reactions or other means.  The role of these manipulations in influencing collision cross sections and therefore equilibration dynamics, is the subject of this chapter.

\section{ Relaxation of ultracold gases taken out of equilibrium }

We begin with a macroscopic perspective and address the collective dynamics of a nonequilibrium molecular gas. 

\subsection{ Molecular dynamics and the Boltzmann equation }

In a nondegenerate gas, the thermal de Broglie wavelength of each gas particle is much smaller than the interparticle spacing, making a classical description of the gas sufficient to encapsulate its many-body physics \cite{Guery99_PRA}. 
In particular, the ensemble phase space distribution $W$ undergoes dynamics governed by the famed Boltzmann transport equation \cite{Reif09_Waveland}: 
\begin{align} \label{eq:Boltzmann_equation}
	\left(
	\frac{ \partial }{ \partial t }
	+
	\frac{ \boldsymbol{p} }{ m }
	\cdot
	\frac{ \partial }{ \partial \boldsymbol{r} }
	-
	\frac{ \partial V(\boldsymbol{r}) }{ \partial \boldsymbol{r} }
	\cdot
	\frac{ \partial }{ \partial \boldsymbol{p} }
	\right)
	W(\boldsymbol{r}, \boldsymbol{p})
	&=
	{\cal I}[W],
\end{align}
where the external confining potential is assumed to be perfectly harmonic $V(\boldsymbol{r}) = (1/2) m \sum_i \omega_i^2 r_i^2$ with trapping frequencies $\omega_i$ along axis $i = x,y,z$, and ${\cal I}[W]$ is the collision integral responsible for two-body scattering events. Functionally, ${\cal I}$ is nonlinear in the phase space distribution:
% \textcolor{red}{[JLB: I'm pretty sure the rest of the world would call this the phase space distribution.]}
\begin{align}
	{\cal I}[W]
	&=
	\int d^3\boldsymbol{p}_1
	| \boldsymbol{p} - \boldsymbol{p}_1 |
	\int d\Omega'
	\frac{ d\sigma }{ d\Omega' }
	( W' W'_1 - W W_1 ),
\end{align}   
because it ignores all two-body quantum coherences generated by each collision, keeping track of only the one-body molecular state and assuming that collisions are perfectly coincident ($\boldsymbol{r}' = \boldsymbol{r}$) \footnote{ The Boltzmann equation can be derived from the Schr\"odinger equation, by performing a Wigner-Weyl transform to the one-body reduced density matrix, and truncating the Mayol bracket to lowest non-trivial order in $\hbar$ \cite{Wigner32_PR, Moyal49_MPCPS, Baker58_PR, Polkovnikov10_AP}. }. 
Above, primes denote post-collision variables and we utilized the commonly adopted short-hand notation $W' = W(\boldsymbol{r}, \boldsymbol{p}')$ and  $W_1 = W(\boldsymbol{r}, \boldsymbol{p}_1)$. 
The post-collision scattering angles $\hat{\boldsymbol{k}}'$, are determined by the differential cross section $\frac{ d \sigma }{ d\Omega }(\hat{\boldsymbol{k}}', \hat{\boldsymbol{k}})$, 
at incident scattering wavevector $\boldsymbol{k} = \hat{\boldsymbol{k}} \sqrt{ 2 \mu E / \hbar^2 }$ with collision energy $E$, between molecules of reduced mass $\mu$ in the center-of-mass frame of reference. 

At thermal equilibrium, the material derivative, and thus the collision integral, must be identically zero, providing us the Maxwell-Boltzmann stationary solution $W_0(\boldsymbol{r}, \boldsymbol{p}) = n_0(\boldsymbol{r}) c_0(\boldsymbol{p})$, with local density $n_0(\boldsymbol{r}) = \exp[ -{ V(\boldsymbol{r}) / (k_B T) } ] / Z_r$ and momentum ensemble $c_0(\boldsymbol{p}) = \exp[ -{ \boldsymbol{p}^2 / (2 m k_B T) } ] / Z_p$, where $Z_r$ and $Z_p$ are the respective partition functions.   
But if brought out of equilibrium, the integro-differential form of Eq.~(\ref{eq:Boltzmann_equation}) makes it difficult to obtain exact solutions, generally requiring numerical methods such as Bird's direct simulation Monte Carlo technique \cite{Bird70_AIP, Bird13_CIPP}. Alternatively, the method of Chapman and Enskog \cite{Chapman90_CUP} has proven highly effective in obtaining approximate analytic solutions in close-to-equilibrium scenarios. 
We describe the latter variety here, both for their simplicity and practical appeal.

Chapman and Enskog found that by tracking only up to second moments of the phase space distribution, a closed set of rate equations can be derived from the Boltzmann equation that describes their dynamics:
\begin{subequations} \label{eq:Enskog_eqns}
\begin{align}
    \dfrac{d \langle r_i^2 \rangle}{d t} 
    - 
    \dfrac{2}{m}\langle r_i p_i \rangle 
    &= 
    0, \label{eq:Enskog_eqns_a} \\
    \dfrac{d \langle p_i^2 \rangle}{d t} 
    +
    2m\omega_i^2\langle r_i p_i \rangle 
    &=
    \mathcal{C}[ p_i^2 ], \\ % \Delta p_i^2
    \dfrac{d \langle r_i p_i \rangle}{dt} 
    -
     \dfrac{1}{m}\left\langle p_i^2 \right\rangle 
    +
    m\omega_i^2 \langle r_i^2 \rangle 
    &=
     0, \label{eq:Enskog_eqns_c}
\end{align}
\end{subequations}
where $\langle \ldots \rangle = (1/N) \int d^3\boldsymbol{r} d^3\boldsymbol{p} W(\boldsymbol{r}, \boldsymbol{p}) ( \ldots )$ denotes an ensemble average, while 
\begin{align}
	{\cal C}[p_i^2]
	&=
	\int \frac{ d^3\boldsymbol{r} }{ N }
    \int \frac{ d^3\boldsymbol{p} d^3\boldsymbol{p}_1 }{ m }
	| \boldsymbol{p} - \boldsymbol{p}_1 |
    W W_1
	\int d\Omega'
	\frac{ d\sigma }{ d\Omega' }
	% ( W' W'_1 - W W_1 )
	% p_i^2,
    \Delta{p_i^2},
\end{align}  
is the phase space averaged collision integral, with collisions resulting in the change $\Delta{p_i^2} = p_i^{\prime 2} + p_{1,i}^{\prime 2} - p_i^{2} - p_{1,i}^{2}$. 
We refer the reader elsewhere for details of deriving Eq.~(\ref{eq:Enskog_eqns}) \cite{Reif09_Waveland, Guery99_PRA, Colussi15_NJP}.
In the absence of collisions, the observables between different axes completely decouple from one another, resulting in conservation of the pseudotemperatures $k_B {\cal T}_i = \langle p_i^2 \rangle/(2 m) + (1/2) m \omega_i^2 \langle r_i^2 \rangle$ (i.e. $d {\cal T}_i / d t = 0$). 
Collisions result in a mixing of energy between the axes, increasing the number of accessible microstates and raising the entropy of the gas until eventual thermalization. 

Motivated by the Maxwell-Boltzmann equilibrium solution, we might expect that the phase space distribution remains close to Gaussian under a small and almost instantaneous perturbation that conserves parity symmetry:
\begin{subequations}
\begin{align}
	W(\boldsymbol{r}, \boldsymbol{p})
	&=
	n(\boldsymbol{r})
	c(\boldsymbol{p}), \\
	n(\boldsymbol{r})
	&\approx
	N \prod_{i=1}^3
	\frac{ 1 }{ \sqrt{ 2\pi \langle r_i^2 
	\rangle } }
	\exp( -\frac{ r_i^2 }{ 2 \langle r_i^2 
	\rangle } ), \\
	c(\boldsymbol{p})
	&\approx
	\prod_{i=1}^3
	\frac{ 1 }{ \sqrt{ 2\pi \langle p_i^2 
	\rangle } }
	\exp( -\frac{ p_i^2 }{ 2 \langle p_i^2 
	\rangle } ).
\end{align}
\end{subequations}
This ansatz reduces the problem to one of following the quadratures in time, which is achieved by the Enskog equations above.
 Besides, any other distribution involving higher moments of $r_i$ and $p_i$ would not have their higher moment dynamics captured by Eq.~(\ref{eq:Enskog_eqns}).  
Adopting the ansatz above, we are able to recast the phase space averaged collision integral in terms of integrals over relative momenta $\boldsymbol{p}_r = \boldsymbol{p} - \boldsymbol{p}_1$ \cite{Reif09_Waveland}:
\begin{align} \label{eq:averaged_collision_integral}
    {\cal C}[p_i^2]
	&=
    \frac{ \langle n \rangle }{ 2 }
    % \frac{1}{N}
    % \int d^3\boldsymbol{r}
    % n(\boldsymbol{r})^2
    \int d^3\boldsymbol{p}_r
    c_r(\boldsymbol{p}_r)
    \frac{ | \boldsymbol{p}_r | }{ 2 m } 
    \int d\Omega'
	\frac{ d\sigma }{ d\Omega' }
    \left( 
    p_{r,i}^{\prime 2} - p_{r,i}^2
    \right),
\end{align}
along with the underlying relative momentum distribution $c_r(\boldsymbol{p}_r)$, that takes the same form of $c(\boldsymbol{p})$ but with the replacement $\boldsymbol{p} \rightarrow \boldsymbol{p}_r$ and all factors of 2 converted to 4. 
It turns out that if the fluctuations of $\langle p_i^2 \rangle$ in $c_r(\boldsymbol{p}_r)$ are treated only up to first-order around their equilibrium value $m k_B T$: 
\begin{align}
    c_r(\boldsymbol{p}_r)
    &\approx 
    c_{r,0}(\boldsymbol{p}_r)
    \left[
    1
    +
    \frac{ 1 }{ 2 }
    \sum_{i=1}^3 
    \left( \frac{ p_i^2 }{ 2 m k_B T } - 1 \right)
    \varepsilon_i 
    \right],
\end{align}
with $\varepsilon_i = \langle p_i^2 \rangle/(m k_B T) - 1$, the integral of Eq.~(\ref{eq:averaged_collision_integral}) can be analytically evaluated for differential cross sections with a finite spherical harmonic expansion.
The resulting functions of $\langle p_i^2 \rangle$ provide a closed set of linear ordinary differential equations for the gas relaxation dynamics, a topic we now turn to.

\subsection{ Rethermalization in the single partial wave regime }

\begin{figure}[ht]
    \centering
    \begin{minipage}{\textwidth}
    \includegraphics[width=\linewidth]{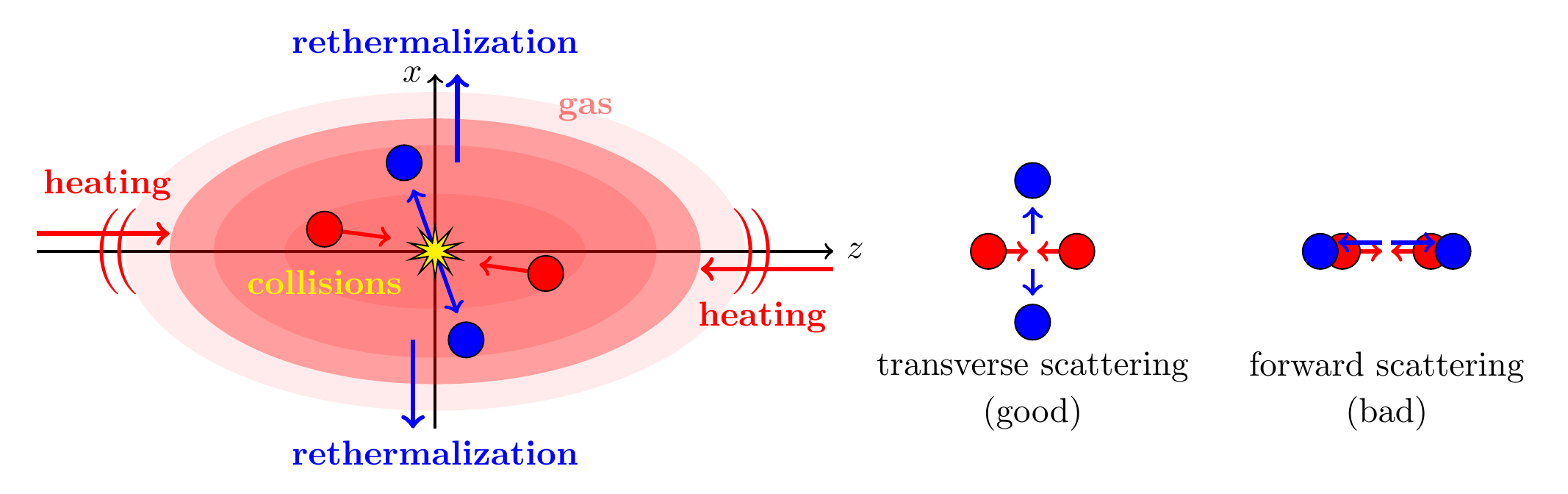}
    \caption{ Cartoon of a cross-dimensional rethermalization experiment. 
    Collisions that scatter primarily into the transverse directions promote thermalization, whereas forward scattering hinders it. 
    Red indicates pre-collision colliders, while post-collision ones are blue. 
    The right two panels illustrate that transverse collisions are ``good" toward rethermalization, while forward scattering is ``bad" as no cross-dimensional mixing occurs.  
    This graphic is taken from Ref.~\cite{Wang24_Boulder}. 
    % \textcolor{red}{[JLB - glad to see it getting out again!]} 
    }
    \label{fig:CDR_cartoon}
    \end{minipage}
\end{figure}

An experimental means to probe the relaxation dynamics of an ultracold gas is to first take it out of equilibrium, then watch it return to a new equilibrium. 
One such scheme to do so is referred to as cross-dimensional rethermalization \cite{Monroe93_PRL}, where the harmonic trapping frequency along one axis is rapidly increased, then held constant as the gas is allowed to re-equilibrate. 
An illustration of this experiment is provided in the leftmost panel of Fig.~\ref{fig:CDR_cartoon}.

A gas that is heated along the $i$-th coordinate will rethermalize along the $j$-th coordinate at a rate $\gamma_{ij}$, following the Boltzmann H-theorem \cite{Reif09_Waveland}. 
When dilute, this rate depends on the density and temperature of the given experiment, whereby it is useful to compare the rethermalization rate to a standard collision rate $\gamma_{\rm coll} = \langle n \rangle \overline{\sigma} v_r$, where $v_r$ is the relative velocity between colliders and $\overline{\sigma} = (4\pi)^{-1} \int d^2\hat{\boldsymbol{k}} \int d^2\hat{\boldsymbol{k}}' d\sigma/d\Omega'$ is the integral scattering cross section. 
Thus for relaxation considerations, the rethermalization rate is proportional to the standard collision rate:
\begin{align} \label{eq:gamma_N_rln}
    \gamma_{ij} 
    &=
    \frac{ \gamma_{\rm coll} }{ {\cal N}_{ij}},
\end{align} 
with proportionality constant ${\cal N}_{ij}$, known as the \textit{number of collisions per rethermalization} \cite{Monroe93_PRL}. 
This constant quantifies the average number of collision instances required before a collision useful for thermalization occurs.  
Then characterizing the relaxation with a single decay rate, this rate can be determined from short-time behavior of the decay and permits a derivation of analytic expressions for ${\cal N}_{i j}$. To formulate this approximation, 
we define the phase space averaged quantity
\begin{align} \label{eq:temperature_differential}
    \langle \chi_i \rangle 
    &=
    k_B ({\cal T}_i - T),
\end{align}
which quantifies the system's deviation from its equilibration temperature $T = T_0 + \delta{\cal T}_i/3$, obtained from the equipartition of energy. $T_0$ denotes the gas temperature before excitation. The relaxation of $\langle \chi_j \rangle$ along axis $j$, following an excitation along axis $i$, is then derived from Eq.~(\ref{eq:Enskog_eqns}) to give differential equation $d \langle \chi_j \rangle = {\cal C}[ \chi_j ] d t$, approximated as
\begin{align} \label{eq:relaxation_time_approx}
    % \frac{\mathcal{C}[ \Delta p_j^2 ]}{2 m},
    {\cal C}[ \chi_j ] 
    \approx 
    - \gamma_{i j} \langle \chi_j \rangle,
\end{align}
in an anisotropic generalization of the relaxation-time approximation \cite{Bohn14_PRA, Reif09_Waveland}, with a decay rate $\gamma_{i j}$. 
The relations in equations (\ref{eq:temperature_differential}) and (\ref{eq:relaxation_time_approx}) above thus identify the rethermalization rate relation
\begin{align} \label{eq:retherm_rate_theory}
    \gamma_{i j}
    =
    - \left. 
    % \frac{ 1 }{ \left( {\cal T}_i(t) - T \right) } 
    \frac{ 1 }{ {\cal T}_j(t) - T } 
    \frac{ d {\cal T}_j(t) }{dt} 
    \right|_{t = 0} 
    \approx 
    -\frac{ {\cal C}[ p_j^2 ] }{ 2 m k_B \left[ {\cal T}_j(0) - T \right] }.
\end{align}
Along with Eq.~(\ref{eq:gamma_N_rln}), the number of collisions per rethermalization can then be cast in terms of the integral \cite{Wang24_PRR}:
\begin{align} \label{eq:Ncoll_integral}
    {\cal N}_{ij} 
    &\approx
    \alpha_{i j} 
    \left(
    \int \frac{ d^3 \boldsymbol{\kappa}  }{ ( 4 \pi )^3 }
    \frac{ e^{ -{ \kappa^2 / 4 } } }{ (4 \pi)^{3/2} }
    \int d^2\Omega'
    \frac{ d\sigma }{ d\Omega' }
    \frac{ \kappa }{ \overline{\sigma} \kappa_{\rm th} }
    \Delta \kappa^2_{i}
    \Delta \kappa^2_{j}
    \right)^{-1},
\end{align}
where $\Delta \kappa_{i}^2 = \kappa_{i}'^2 - \kappa_{i}^2$ is the collisional change in adimensional relative momenta $\boldsymbol{\kappa} = \boldsymbol{p}_{r} / \sqrt{ m k_B T }$, $\kappa_{\rm th} = 4/\sqrt{\pi}$ is the adimensional mean collision velocity, and the excitation-rethermalization geometric factor is defined by $\alpha_{i j} = 2/3$ if $i = j$, otherwise $\alpha_{i j} = -1/3$. The geometric factor $\alpha_{i j}$ is a result of the differential temperature definition of $\langle \chi_i \rangle$ (\ref{eq:temperature_differential}), and the equipartition of thermal energy.

At ultracold temperatures, scattering off finite-range potentials that fall off with intermolecular distance $r$, faster than $r^{-3}$, are known to involve a small number of collisional partial waves $\ket{L, m_L}$ \cite{Sadeghpour00_JPB}.
For isotropic interaction potentials, these partial waves remain uncoupled during a collision so that only one dominates the scattering behavior at low energies $k a_{\rm range} \ll 1$, with $a_{\rm range}$ being the effective range of interactions and $k$ the collision wavenumber. 
Then by virtue of exchange symmetry between indistinguishable and identical particles, low energy collisions of bosons are $L=0$ partial wave (i.e. $s$-wave) dominated, while fermions have predominantly $L=1$ partial wave (i.e. $p$-wave) collisions.
The close-to-threshold scattering phase shifts \cite{Mott65_CP} for each of these partial waves are given by
\begin{subequations}
\begin{align}
	\delta_s(k) 
	=
	-k a_s, \\
	\delta_p(k) 
	=
	-\frac{1}{3}
	k^3 V_p, 
\end{align}
\end{subequations}
where $a_s$ is the $s$-wave scattering length and $V_p$ is the $p$-wave scattering volume. 
The corresponding differential cross sections are then
\begin{subequations}
\begin{align}
	\frac{ d \sigma_s }{ d\Omega }
	&=
	% \frac{ 8\pi^2 }{ k^2 }
	% \abs{ 
	% \frac{ 1 - e^{2 i \delta_s} }{ 4 \pi } 
 %    }^2 
	% \approx
	2 a_s^2, \\
	\frac{ d \sigma_p }{ d\Omega }
	&=
	% \frac{ 2 }{ k^2 }
	% % \abs{ \sum_{m_{\ell}=-1}^{+1} \abs{ Y_{1, m_{\ell}}(\theta, \phi) }^2 
	% % \left( 1 - e^{2 i \delta_p} \right) }^2 
 %    \abs{ 
 %    3 \big( 
 %    \hat{\boldsymbol{k}} \cdot \hat{\boldsymbol{k}}' 
 %    \big) % \cos\theta  
 %    e^{i \delta_p} \sin\delta_p 
 %    }^2 
	% %	\nonumber\\
	% %	&=	
	% %	\frac{ 8\pi }{ k^2 }	
	% %	\abs{ \frac{2}{3} k^3 V_p \sum_{m_{\ell}=-1}^{+1} \abs{ Y_{1, m_{\ell}}(\theta, \phi) }^2 }^2 \nonumber\\
	% \approx
	2
    \big( 
    \hat{\boldsymbol{k}} \cdot \hat{\boldsymbol{k}}' 
    \big)^2
    k^4 V_p^2,
\end{align}
\end{subequations}
which imply the respective integral cross sections $\sigma_s = 8\pi a_s^2$ and $\sigma_p = (8/3)\pi k^4 V_p^2$. 
Plugging these formulas into Eq.~(\ref{eq:Ncoll_integral}) grants us the rethermalization coefficients
\begin{subequations}
\begin{align}
    {\cal N}_s 
    &=
    { 5 / 2 }, \\
    {\cal N}_p
    &=
    { 25 / 6 },
\end{align}
\end{subequations}
that agree with Monte Carlo simulations \cite{DeMarco99_PRL}.

\subsection{ Anisotropic relaxation of oriented dipoles \label{sec:DDI_rethermalization} }

The molecular collisions of interest here are, however, not sufficiently handled by just a single $s$ or $p$ partial wave, but formally involve all partial waves for a complete description, a consequence of the $r^{-3}$ long-range behavior of the dipole-dipole interaction \cite{Bohn14_PRA}. 
The innate molecular-frame dipole moments in bialkali molecules would cause pairs of them, labeled molecule $A$ and $B$, to interact at long-range via dipole-dipole interactions:
\begin{align} \label{eq:DDI_potential}
	V_{\rm dd}(\boldsymbol{r})
	&=
	\frac{ \boldsymbol{d}_A \cdot \boldsymbol{d}_B - 3 ( \boldsymbol{d}_A \cdot \hat{\boldsymbol{r}} ) ( \boldsymbol{d}_B \cdot \hat{\boldsymbol{r}} ) }{ 4 \pi \epsilon_0 r^3 },
\end{align}
where $\boldsymbol{d}_{\alpha}$ is the electric dipole moment of molecule $\alpha = A,B$, and $\epsilon_0$ is the vacuum electric constant. 
These dipole moments can be oriented by external electric fields \cite{Bohn09_CRC}, for which if appropriate to induce a well-defined dipole orientation, results in a first-order point dipole interaction
\begin{align} \label{eq:point_DDI}
    V_{\rm dd}(\boldsymbol{r})
	&\approx
	\frac{ d_{\rm eff}^2 }{ 4 \pi \epsilon_0 }
    \frac{ 1 - 3 \cos^2\theta }{ r^3 },
\end{align}
where $\theta$ is the angle between $\boldsymbol{r}$ and the external field axis which the dipoles are oriented along $\hat{\boldsymbol{{\cal E}}} = \hat{\boldsymbol{d}}_{\rm eff}$, and $d_{\rm eff}$ is the effective lab-frame dipole moment along $\hat{\boldsymbol{{\cal E}}}$. 
We will return to a more explicit treatment of inducing these first-order dipole-dipole interactions later, but take it at face value for now.

\begin{figure}[ht]
    \centering
    \begin{minipage}{\textwidth}
    \centering
    \includegraphics[width=0.8\linewidth]{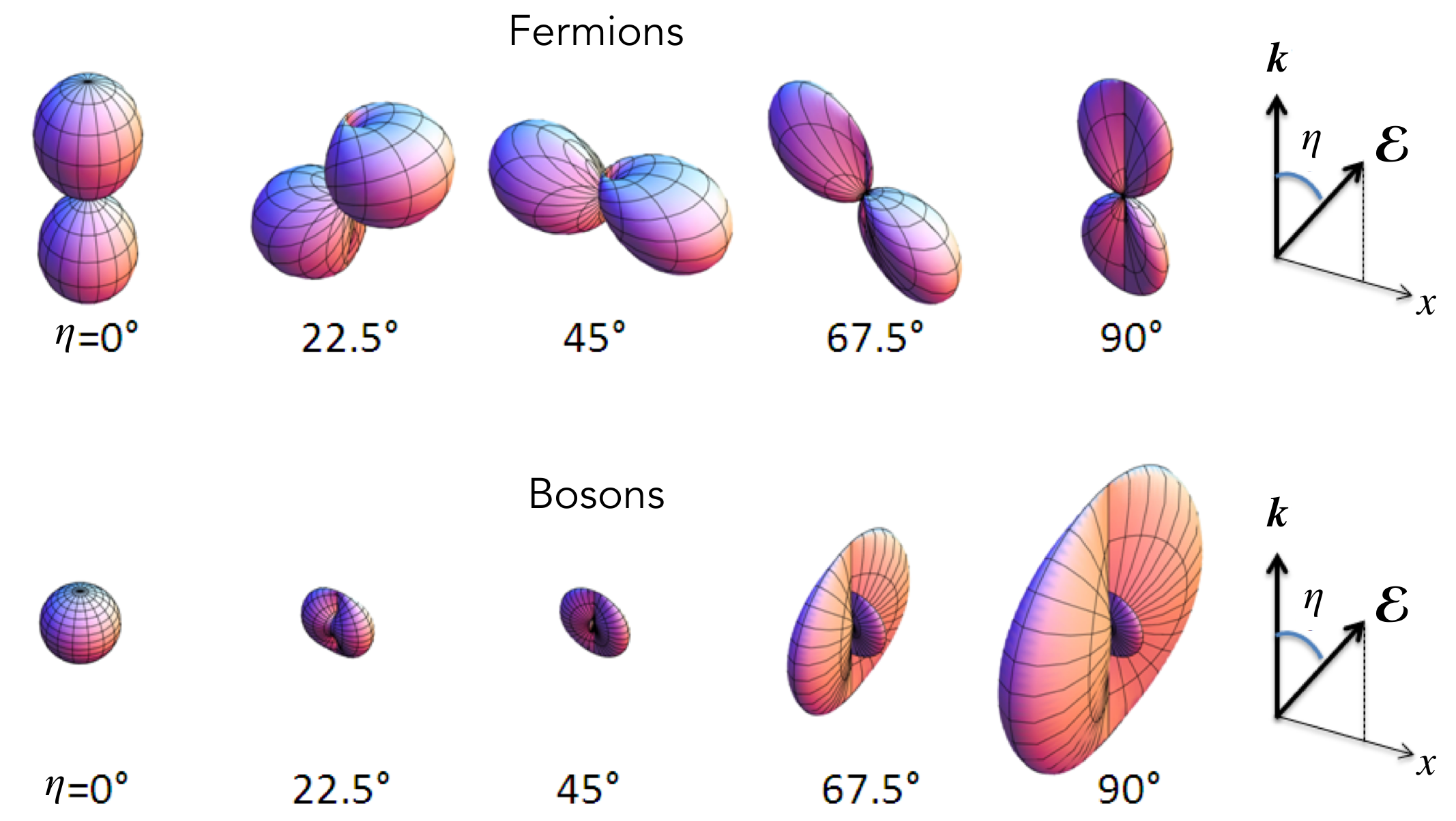}
    \caption{ The anisotropic differential cross sections (in arbitrary units) between point dipolar particles for various deflection angles $\eta = \cos^{-1}\hat{\boldsymbol{k}} \cdot \hat{\boldsymbol{{\cal E}}}$, comparing identical fermions (upper panel) and identical bosons (lower panel). This plot is reproduced with data from Ref.~\cite{Bohn14_PRA}. }
    \label{fig:differential_scattering}
    \end{minipage}
\end{figure}

The anisotropy of dipole-dipole interactions can favor the scattering of particles into specific directions in space, leading to effects observable not only at the two-body level, but also macroscopically in the nonequilibrium gas dynamics. 
The number of collisions per rethermalization, defined above, has been exhaustively explored for particles undergoing purely dipolar interactions \cite{Bohn14_PRA, Aikawa14_PRL, Wang21_PRA}, including isotropic s-wave scattering in the case of identical bosons \cite{Tang15_PRA, Patscheider22_PRA}.
For sufficiently low temperatures, ultracold collisions are governed by their threshold laws, for which scattering off a $r^{-3}$ potential approaches an energy independent cross section \cite{Bohn09_NJP}. 
For scattering with nonzero partial waves, the angular momentum barrier prevents low energy scatterers from seeing much else other than the long-range tail of the dipolar $r^{-3}$ potential before being reflected. 
As a result, the dipole-dipole interactions only weakly perturb the incident pair molecular wavefunction, allowing use of the Born approximation to compute scattering quantities. 
For instance, the scattering amplitude up to first-order in Born approximation evaluates to \cite{Bohn14_PRA}:
\begin{align} \label{eq:Born_scattering_amplitude}
    f( \hat{\boldsymbol{k}}', \hat{\boldsymbol{k}} )
    &\approx
    -a_s
    -
    \frac{ 2 \mu }{ \hbar^2 }
    \frac{ 1 }{ 4 \pi }
    \int d^3 \boldsymbol{r}
    e^{ -i \boldsymbol{k}' \cdot \boldsymbol{r} }
    V_{\rm dd}( \boldsymbol{r} )
    e^{ i \boldsymbol{k} \cdot \boldsymbol{r} } \nonumber\\
    &=
    -a_s
    +
    a_d
    \Bigg(
    \frac{ 2 }{ 3 }
    -
    \frac{ ( \hat{\boldsymbol{k}} \cdot \hat{\boldsymbol{{\rm E}}}_{\rm d.c.}
    -
    \hat{\boldsymbol{k}}' \cdot \hat{\boldsymbol{{\rm E}}}_{\rm d.c.} )^2 }{ 1 - \hat{\boldsymbol{k}} \cdot \hat{\boldsymbol{k}}' }
    \Bigg), 
\end{align}
where $a_s$ is the $s$-wave scattering length and $a_d = \mu d_{\rm eff}^2 / (4 \pi \epsilon_0 \hbar^2)$ is the dipole length. 
For collisions of identical indistinguishable particles, the scattering amplitude must be appropriately symmetrized according to $f_{F, B}(\hat{\boldsymbol{k}}, \hat{\boldsymbol{k}}') = [ f(\hat{\boldsymbol{k}}, \hat{\boldsymbol{k}}') \pm f(\hat{\boldsymbol{k}}, -\hat{\boldsymbol{k}}') ] / \sqrt{ 2 }$, with subscript $B$ denoting identical bosons (+), and $F$ fermions (-).
From these scattering amplitudes, we then have access to analytic forms for the differential cross sections through the relation 
\begin{align} \label{eq:DCS_scatAmp2}
    \frac{ d\sigma_{F,B} }{ d\Omega }
    (\hat{\boldsymbol{k}}, \hat{\boldsymbol{k}}')
    &=
    \big| f_{F,B}(\hat{\boldsymbol{k}}, \hat{\boldsymbol{k}}') \big|^2.
\end{align}
Because of the non-trivial angular dependence in $V_{\rm dd}$, the differential cross sections inherit significant anisotropies illustrated by three-dimensional surface plots of $d\sigma/d\Omega$ in Fig.~\ref{fig:differential_scattering}, as a function of the direction of the outbound scattering wave vector $\hat{\boldsymbol{k}}'$. Each surface plot has the dipoles oriented along a different tilt angle $\eta = \cos^{-1}\hat{\boldsymbol{k}} \cdot \hat{\boldsymbol{{\cal E}}}$, between the incident direction $\hat{\boldsymbol{k}} = \hat{\boldsymbol{z}}$ and the dipole direction set, for instance, by an applied d.c. electric field ${\boldsymbol{{\rm E}}}_{\rm d.c.}$.
The upper and lower panels compare the differential cross sections for identical dipolar fermions and bosons respectively.

\begin{figure}[ht]
    \centering
    \begin{minipage}{\textwidth}
    \centering
    \includegraphics[width=\linewidth]{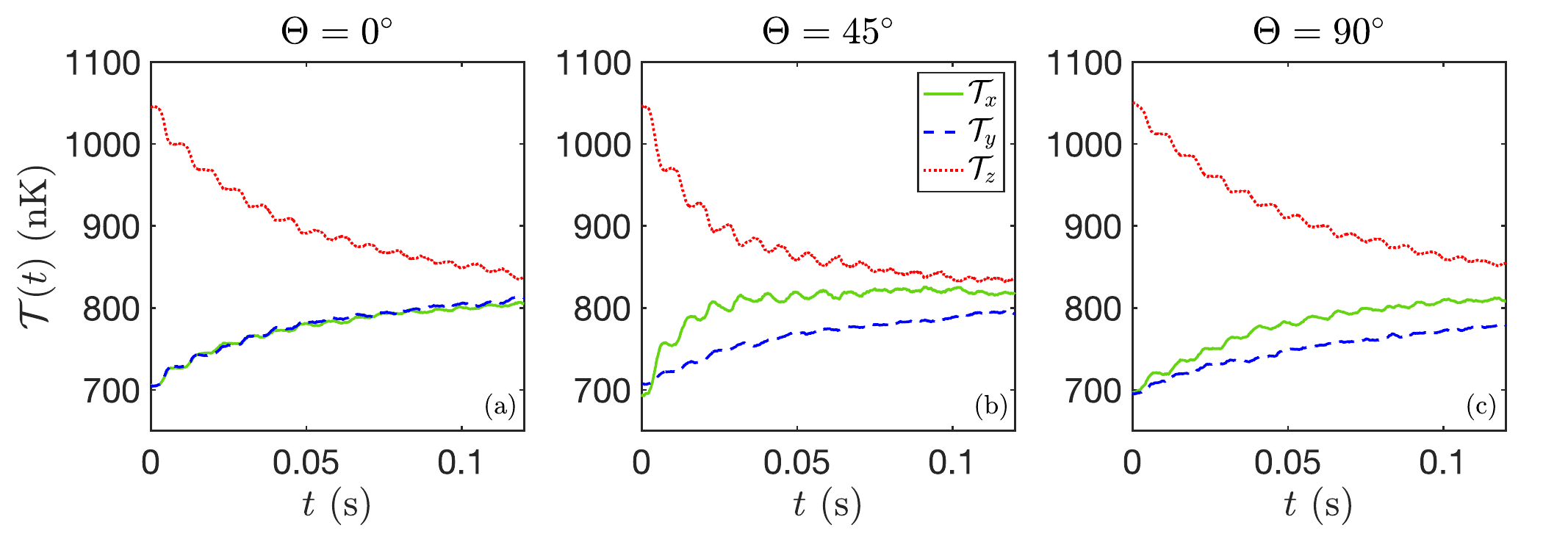}
    \caption{ Numerically simulated pseudotemperature traces ${\cal T}_x(t)$ (solid green curves), ${\cal T}_y(t)$ (dashed blue curves) and ${\cal T}_z(t)$ (dotted red curves) for 3 values of $\Theta = 0^{\circ}, 45^{\circ}, 90^{\circ}$, in subplots (a), (b) and (c) respectively. The simulations assume a gas of 2000 $^{23}$Na$^{40}$K molecules with an effective dipole moment of $d_{\rm eff} = 0.75$ D, in a harmonic trap of mean trap frequency $\overline{\omega} = 2\pi \times 100$ Hz. This plot is reproduced with data from Ref.~\cite{Wang23_PRA}.  }
    \label{fig:NaK_diluteCDR}
    \end{minipage}
\end{figure}

With anisotropic cross sections, dipolar molecules preferentially transfer momentum into certain directions, relative to $\hat{\boldsymbol{d}}$, over others during a collision.  
Consequently, the collective dynamics of a collisional molecular gas will reflect these directional biases when taken out of equilibrium, an effect distinctly observable in cross-dimensional rethermalization experiments. 
Tracking the nonequilibrium pseudotemperatures over time, a marked difference in the rethermalization rates between ${\cal T}_x$ and ${\cal T}_y$ is observed in Fig.~\ref{fig:NaK_diluteCDR}, as the dipoles are tilted away from $\Theta = \cos^{-1} \hat{\boldsymbol{z}} \cdot \hat{\boldsymbol{d}}_{\rm eff} = 0$, to break cylindrical symmetry along $z$. The angle $\Theta$ is defined with the dipoles assumed to lie in the $x,z$-plane.
These plots are obtained from numerical simulations of $^{23}$Na$^{40}$K molecules with the direct simulation Monte Carlo method \cite{Bird70_AIP}, proven to produce dynamics consistent with experimental observations \cite{Tang15_PRA, Patscheider22_PRA, Wang24_PRA2}.

More generally, dilute gases of indistinguishable dipolar fermions will have their thermalization characterized by ${\cal N}_{ij}(\Theta)$, which take the concise analytic forms \cite{Wang21_PRA}:
\begin{subequations} \label{eq:NCPR_fermions}
\begin{align}
    {\cal N}_{xx}(\Theta) 
    &=
    \frac{112}{45 - 4 \cos(2\Theta) - 17 \cos(4\Theta)}, \\
    {\cal N}_{yx}(\Theta) 
    &= 
    \frac{14}{3 - \cos(2\Theta)}, \\
    {\cal N}_{zx}(\Theta) 
    &= 
    \frac{56}{33 - 17 \cos (4 \Theta )}, \\
    {\cal N}_{yy}(\Theta) 
    &= 
    \frac{14}{3}, \\
    {\cal N}_{zy}(\Theta) 
    &= 
    \frac{14}{3 + \cos(2\Theta)}, \\
    {\cal N}_{zz}(\Theta) 
    &= 
    \frac{112}{45 + 4 \cos(2\Theta) - 17 \cos(4\Theta)}.
\end{align}
\end{subequations}
Similar formulas can also be derived for indistinguishable dipolar bosons, although involvement of the $s$-wave scattering length complicates these formulas, so we refer the reader elsewhere for these coefficients \cite{Wang21_PRA}.

\begin{figure}[ht]
    \centering
    \begin{minipage}{\textwidth}
    \centering
    \includegraphics[width=\linewidth]{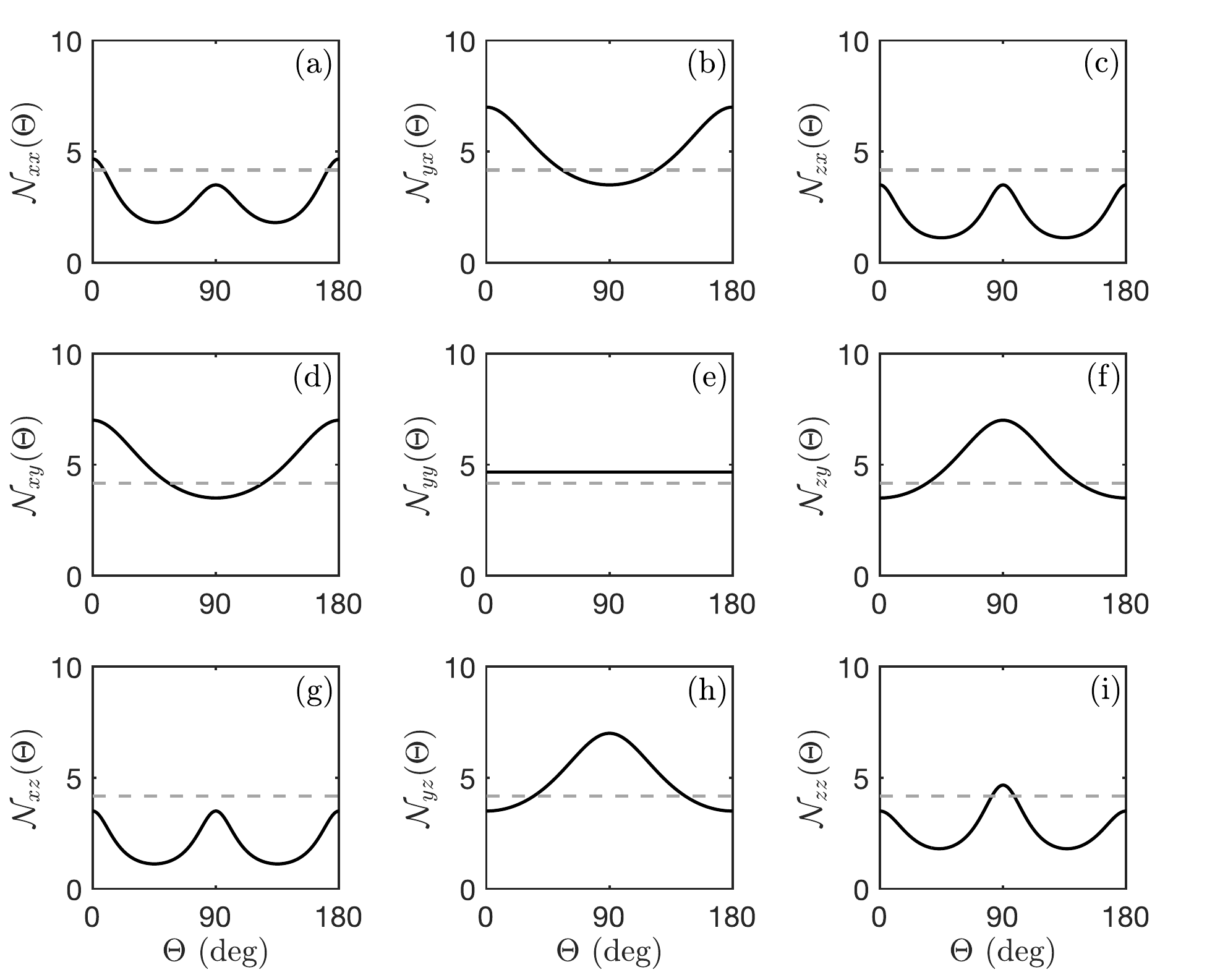}
    \caption{ The number of collisions per rethermalization for identical dipolar fermions ${\cal N}_{i j}$ (solid black curves) as a function of $\Theta$, for all nine excitation-rethermalization configurations from Eq.~(\ref{eq:NCPR_fermions}). For comparison, we also plot ${\cal N}_p = 25/6$ for $p$-wave scattering (dashed gray lines). This plot is reproduced with data from Ref.~\cite{Wang21_PRA}. }
    \label{fig:NCPR_fermions}
    \end{minipage}
\end{figure}

\section{ Not just point dipoles: collisional shielding of molecules }

Thus far, we have been rather flippant % \textcolor{red}{(we're so naughty!)} 
about the exact form of native molecule-molecule interactions, assuming them to be that between two structureless point dipoles. In laboratory experiments, the magnetic analog of these interaction potentials is actually quite reasonable for understanding ultracold gases of paramagnetic atoms such as Cr \cite{Griesmaier05_PRL}, Dy \cite{Lu11_PRL}, Er \cite{Aikawa12_PRL}, or Eu \cite{Miyazawa22_PRL}. However for polar molecules, such idealized interactions only occur in rather fine-tuned settings, generally involving quantum state preparation, then applying external electromagnetic fields to tune specific pairwise molecular transitions into resonance \cite{Avdeenkov06_PRA, Quemener16_PRA, GonzalezMartinez17_PRA, Karman18_PRL, Lassabliere18_PRL}.  It turns out that a proper treatment of these interactions is absolutely necessary in ultracold collisions of bialkali molecules, due to the nigh-inevitable exothermic loss that awaits them if they enter into close proximity (typically $\lesssim 10$ nm) with one another.

Albeit generally non-trivial to implement, the benefits of engineering intermolecular interactions with external fields are two-fold: (1) the external field parameters can be used as tuning knobs to scale the strength of intermolecular interactions, (2) and the literal shape of interaction potential surfaces can be manipulated to vary the anisotropy, and minimum distances over which the molecules are allowed to interact. 
Achieving a collisionally stable, dipolar interacting gas of molecules requires leveraging both these capabilities. 
Key to achieving such quantum control over collisions is an ultracold environment. 
At temperatures of $\sim 100$ nK ($\sim$ kHz), preparing molecules in their ground electronic and vibrational state makes collisional excitation of these degrees of freedom energetically inaccessible. The hyperfine structure can also be made a spectator by applying a strong static magnetic field to induce large Zeeman splittings, allowing single spin state addressability and suppression of hyperfine transitions. 
The result is an effective reduction of the entire molecular Hilbert space to just rigid-body rotations, affected by electrostatic interactions that apply mechanical torques as two molecules approach.  
In the absence of any external electric fields, dipole-dipole interactions are experienced only as a second-order effect, resulting in dispersive $\sim r^{-6}$ interactions. 
First-order dipole-dipole interactions between polar molecules is only realized by dressing with external static electric or microwave fields, as will be the focus of the subsequent sections.

\subsection{ Collisional shielding with a static electric field }

Molecules, somewhat notoriously, have the tendency to undergo exothermic transitions (chemical reactions) when they meet at close enough proximities \cite{Mayle12_PRA, Ni10_Nat, Bause23_JPCA}.  
Such an event would be disastrous for an ultracold gas, unless studying these transitions were the point.  For applications of ultracold molecules to many-body physics \cite{Carr09_IOP}, it is convenient to engineer potential energy surfaces between molecules that are repulsive at some suitable intermolecular distance (often hundreds of Bohr radii), thus shielding the molecules from getting too close. The idea of collisional shielding is predicated on engineering interaction-induced avoided crossings, by careful tuning of the noninteracting molecular spectra with externally applied fields.  

\begin{figure}[ht]
    \centering
    \begin{minipage}{\textwidth}
    \centering
    \includegraphics[width=1.0\columnwidth]{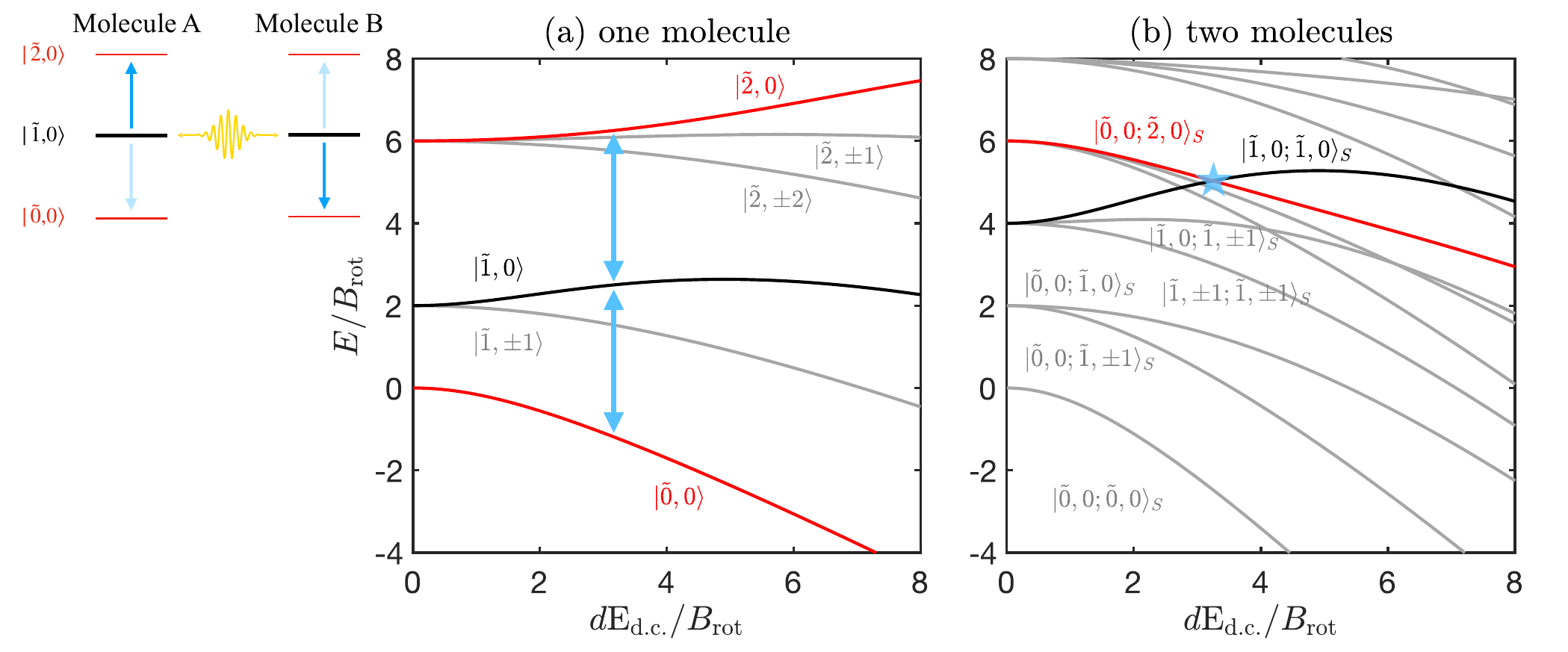}
    \caption{ The (a) one and (b) two-molecule rigid rotor spectra as a function of d.c. electric field. The blue arrows in (a) indicate the resonant transitions in each molecule during F\"orster resonant shielding, while the blue star in (b) shows the combined molecular state crossing, that becomes avoided in the presence of dipole-dipole interactions. Not all states have been labeled to avoid clutter in the plot. The top left inset shows a schematic for the resonant energy exchange process between the two colliding molecules.   
    See the main text for more details. }
    \label{fig:one_n_two_moleculeSpectrum}
    \end{minipage}
\end{figure}

This section will explain the key ideas of collisional shielding for the case of a static electric field  $\boldsymbol{{\rm E}}_{\rm d.c.}$. 
For simplicity we will treat the diatomic molecules of concern here simply as rigid rotors, which is often times a good enough approximation.    
Then for a single rigid-rotor molecule in a static electric field, the Hamiltonian involves the intrinsic molecular rotational energy and coupling of its electric dipole to the field:
\begin{align}
    {\cal H}
    &=
    {\cal H}_{\rm rot} 
    +
    {\cal H}_{\rm d.c.} 
    =
    B_{\rm rot} \boldsymbol{N}^2
    -
    \boldsymbol{d} \cdot \boldsymbol{{\rm E}}_{\rm d.c.},
\end{align}
where $\boldsymbol{N}$ is the rotor rotational angular momentum vector. Utilizing the rotor basis $\ket{ N, m_N }$ that diagonalizes ${\cal H}_{\rm rot}$, the matrix elements of ${\cal H}_{\rm d.c.}$ are 
\begin{align}
    \bra{ N', m'_N }
    \mathcal{H}_{\text{d.c.}} 
    \ket{ N, m_N }
    &=
    -d {\rm E}_{\rm d.c.} ( -1 )^{m_N}
    \sqrt{ (2 N + 1) (2 N' + 1) } \nonumber\\
    &\quad\quad\quad\quad \times  
    \begin{pmatrix}
        N' & 1 & N \\
        0 & 0 & 0
    \end{pmatrix}
    \begin{pmatrix}
        N' & 1 & N \\
        -m'_N & 0 & m_N
    \end{pmatrix}.
\end{align} 
Diagonalization of the molecule-field Hamiltonian ${\cal H}$, provides the field-dressed eigenstates $|{ \tilde{N}, m_{\tilde{N}} }\rangle$ and eigenenergies $\epsilon_{\tilde{N}, m_{\tilde{N}}}$ plotted for several low-lying rotational states in Fig.~\ref{fig:one_n_two_moleculeSpectrum}\textcolor{blue}{a}. 
Tildes on $\tilde{N}$ denote d.c. field-dressed quantum numbers, although $m_{\tilde{N}}$ remains a good quantum number so we omit tildes on it.  
Although zero in the absence of the field, the field-dressed eigenstates now develop permanent dipole moments in the laboratory frame, along with modified transition dipole moments between various field-dressed states:
\begin{align} \label{eq:DC_dipole_moment}
    d^{(m_{\tilde{N}})}_{ \tilde{N} \rightarrow \tilde{N}' }
    &=
    \langle{ \tilde{N}', m_{\tilde{N}} }|
    \boldsymbol{d} \cdot \hat{\boldsymbol{{\rm E}}}_{\rm d.c.}
    |{ \tilde{N}, m_{\tilde{N}} }\rangle 
    =
    -% \left. 
    \frac{ \partial \epsilon_{\tilde{N}, m_{\tilde{N}}} }{ \partial {\rm E}_{\rm d.c.} },
    % \right|_{{\rm E}_{\rm d.c.}^*}, 
\end{align}
permitting first-order dipole-dipole interactions.  

\subsubsection{ Simplified model and effective potential }

\begin{figure}[ht]
    \centering
    \begin{minipage}{\textwidth}
    \centering
    \includegraphics[width=\linewidth]{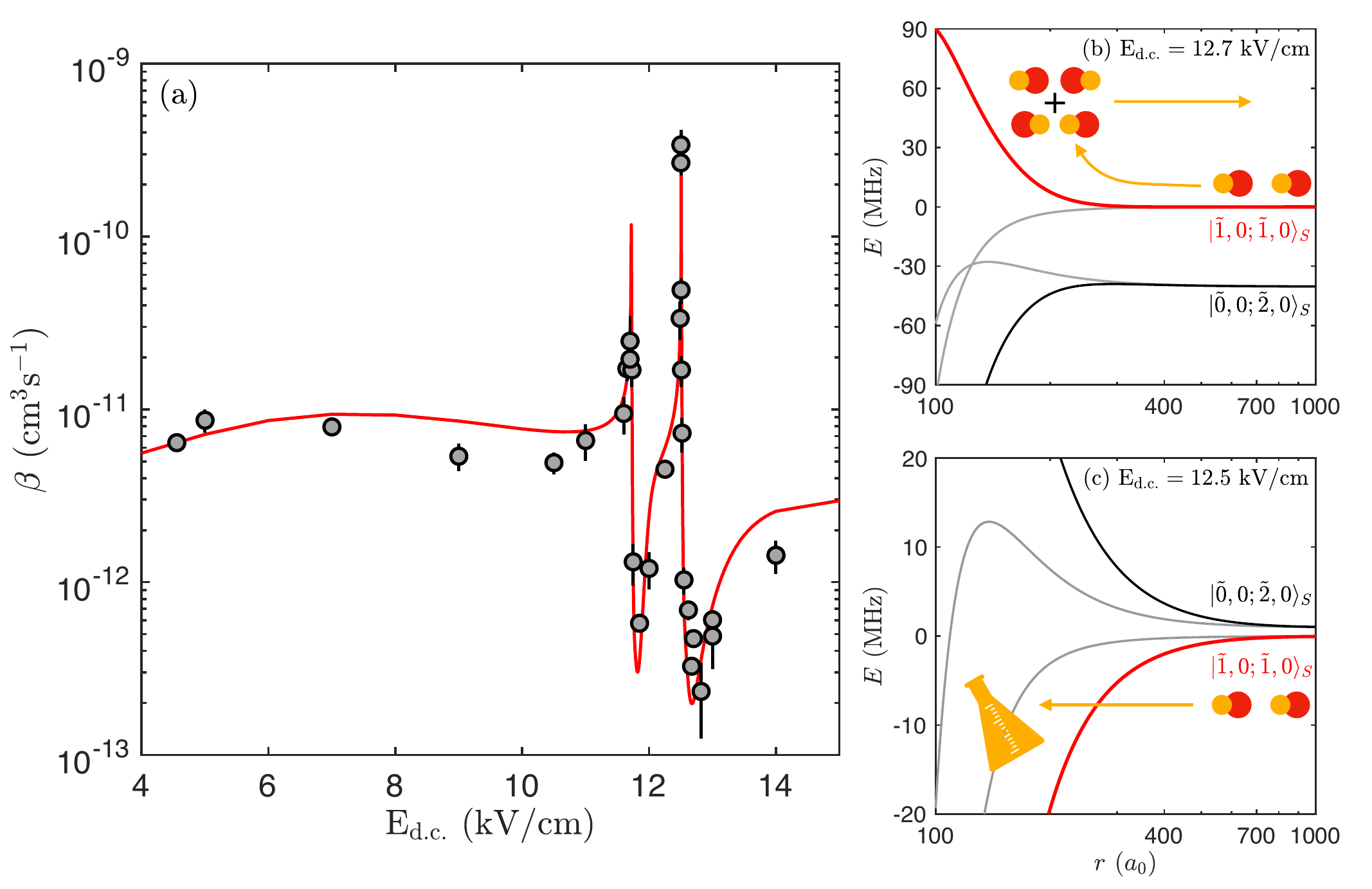}
    \caption{ (a) Collisional quenching rate coefficient $\beta$ as a function of applied static electric field for an ultracold gas of KRb molecules, from experimental measurements (points with error bars) and close-coupling calculations (solid red line). This subplot is reproduced with data from Ref.~\cite{Li21_NatPhys}. 
    Potential energy curves for head-on collisions ($\theta = 0$) as a function of intermolecular distance for pair states involved in the F\"orster resonance, at the ${\rm E}_{\rm d.c.} = 12.7$ kV/cm shielding field (b), and slightly away from the shielding resonance at ${\rm E}_{\rm d.c.} = 12.5$ kV/cm (c). The illustration in subplot (b) indicates an effective reorientation of the molecules that repels them from the short-range as they stay on the shielded adiabat (upper red curve), while subplot (c) shows the molecules entering the short-range on the unshielded adiabat (lower red curve) where chemistry occurs. }
    \label{fig:KRb_DCshielding}
    \end{minipage}
\end{figure}

% \textcolor{blue}{Add words about static field shielding.}
As the electric field is varied, the joint rotational spectrum of two molecules $A$ and $B$, experiences level crossings, one of which is indicated by the blue star in Fig.~\ref{fig:one_n_two_moleculeSpectrum}\textcolor{blue}{b} between the dressed states $| \tilde{N}_A, m_{\tilde{N}_A} \rangle$ $| \tilde{N}_B, m_{\tilde{N}_B} \rangle = | \tilde{1}, 0 \rangle| \tilde{1}, 0 \rangle$ and $| \tilde{0}, 0 \rangle| \tilde{2}, 0 \rangle$. 
Equivalently, this crossing is a result of the $| \tilde{0}, 0 \rangle \rightarrow | \tilde{1}, 0 \rangle$ and $| \tilde{1}, 0 \rangle \rightarrow | \tilde{2}, 0 \rangle$ transitions being tuned into resonance by the d.c. Stark effect, illustrated by the blue arrows in Fig.~\ref{fig:one_n_two_moleculeSpectrum}\textcolor{blue}{a}.
If the molecules were now to approach close to this resonant field strength, dipole-dipole interactions would couple these nearly degenerate molecular states, resulting in F\"orster resonant energy exchange \cite{Foerster46_Nat, Foerster49_ZN, Lassabliere22_PRA} or an avoided crossing in the time-independent picture. 
By preparing the molecules in their $| \tilde{N}, m_{\tilde{N}} \rangle = | \tilde{1}, 0 \rangle$ rotational states, tuning the d.c. electric field strength across these resonances results in a variation of the collisional loss rate $\beta$ by several orders of magnitude, showcased for ultracold KRb molecules in Fig.~\ref{fig:KRb_DCshielding}\textcolor{blue}{a}. 
In this plot, experimental measurements (circles with error bars) from the JILA KRb group are compared to numerical scattering calculations (solid red curve) performed with the coupled-channel formalism \cite{Quemener17_RSC}. The excellent agreement assures us that the theoretical models we present here provide a strong basis of understanding to the actual physics of shielding.

In general, a full description of molecular scattering in static electric fields can be rather involved, requiring treatment of the strong dipole-dipole couplings between bare rotational states \cite{Quemener17_RSC}:
\begin{align} \label{eq:DDI_matrix_elements}
    % \bra{ L', m'_L }
    \langle N'_A, m'_{N_A}; & N'_B, m'_{N_B} |
    V_{\rm dd}(\boldsymbol{r})
    | N_A, m_{\tilde{N}_A}; N_B, m_{\tilde{N}_B} \rangle
    % \ket{ L, m_L } 
    \nonumber\\
    &=
    % -\frac{ d^2 }{ 4 \pi \epsilon_0 r^3 }
    % % \bra{ L', m'_L } 
    % C_{2,-q}(\theta, \phi)
    % % \ket{ L, m_L}
    % \sqrt{ 30 } 
    -\frac{ C_{2,-q}(\theta, \phi) }{ 4 \pi \epsilon_0 r^3 } 
    \sqrt{ 30 } (-1)^{m_{N_A} + m_{N_B}}
    \begin{pmatrix}
        1 & 1 & 2 \\
        q_A & q_B & -q
    \end{pmatrix} \nonumber\\
    &\quad \times
    d \sqrt{ (2 N'_A + 1) (2 {N_A} + 1) }
    \begin{pmatrix}
        N'_A & 1 & {N_A} \\
        -m'_{N_A} & q_A & m_{N_A}
    \end{pmatrix}
    \begin{pmatrix}
        N'_A & 1 & {N_A} \\
        0 & 0 & 0
    \end{pmatrix} \nonumber\\
    &\quad \times
    d \sqrt{ (2 N'_B + 1) (2 {N_B} + 1) }
    \begin{pmatrix}
        N'_B & 1 & {N_B} \\
        -m'_{N_B} & q_B & m_{N_B}
    \end{pmatrix}
    \begin{pmatrix}
        N'_B & 1 & {N_B} \\
        0 & 0 & 0
    \end{pmatrix},
\end{align}
where $q_A = m'_{N_A} - m_{{N}_A}$, $q_B = m'_{N_B} - m_{{N}_B}$ and $q = q_A + q_B$ are the angular momentum transfers, and 
\begin{align}
    C_{\ell, m_{\ell}}(\theta, \phi)
    &=
    \sqrt{ \frac{ 4 \pi }{ 2\ell + 1 } }
    Y_{\ell, m_{\ell}}(\theta, \phi),
\end{align}
are reduced spherical harmonics. 
However, because shielding mainly relies on the avoided crossing between the combined two-molecule states $| \tilde{N}_A, m_{\tilde{N}_A}; \tilde{N}_B, m_{\tilde{N}_B} \rangle_S = | \tilde{1}, 0; \tilde{1}, 0 \rangle_S$ and $| \tilde{0}, 0; \tilde{2}, 0 \rangle_S$, much insight can be gleaned from a reduced model including just these two states.  
The subscript $S$ on the states denote that they have been appropriately symmetrized under particle exchange. 
In this truncated basis, the interactions result in a pair of adiabatic potential energy surfaces, one principally repulsive and one principally attractive.  
These adiabats are illustrated by the solid black and red curves in subplots (b) and (c) of Fig.~\ref{fig:KRb_DCshielding}.
Shielding is achieved when the collisional entrance channel $| \tilde{1}, 0; \tilde{1}, 0 \rangle_S$ sits above $| \tilde{0}, 0; \tilde{2}, 0 \rangle_S$, showcased in subplot (b), allowing the diabatic crossing (solid gray curves) to be avoided by dipole-dipole coupling. 
The upper repulsive adiabat is then well approximated by \cite{Lassabliere22_PRA}:
\begin{align} \label{eq:shielded_adiabat}
    V_{\rm eff}(r, \theta)
    &=
    \frac{ 1 }{ 2 } ( V_a + V_b ) 
    +
    \frac{ 1 }{ 2 } \sqrt{ ( V_a - V_b )^2 + 4 W^2 },
\end{align}
constructed out of the diabatic pieces  
\begin{subequations}
\begin{align}
    V_a(r, \theta)
    &=
    % \frac{ \chi_a }{ 4 \pi \epsilon_0 r^3 } ( 1 - 3 \cos^2\theta )
    % -\frac{ \chi_a }{ \pi \epsilon_0 r^3 } \sqrt{ \frac{ \pi }{ 5 } } Y_{2,0}(\theta)
    -\frac{ d^{(0)}_{\tilde{1} \rightarrow \tilde{1}} d^{(0)}_{\tilde{1} \rightarrow \tilde{1}} }{ \pi \epsilon_0 r^3 } \sqrt{ \frac{ \pi }{ 5 } } Y_{2,0}(\theta)
    +
    E_{\tilde{1} \tilde{1}}, \\
    V_b(r, \theta)
    &=
    % \frac{ \chi_b }{ 4 \pi \epsilon_0 r^3 } ( 1 - 3 \cos^2\theta )
    % -\frac{ \chi_b }{ \pi \epsilon_0 r^3 } \sqrt{ \frac{ \pi }{ 5 } } Y_{2,0}(\theta)
    -\frac{ d^{(0)}_{\tilde{0} \rightarrow \tilde{0}} d^{(0)}_{\tilde{2} \rightarrow \tilde{2}}
    +
    d^{(0)}_{\tilde{0} \rightarrow \tilde{2}} d^{(0)}_{\tilde{2} \rightarrow \tilde{0}} }{ \pi \epsilon_0 r^3 } \sqrt{ \frac{ \pi }{ 5 } } Y_{2,0}(\theta)
    +
    E_{\tilde{0} \tilde{2}}, \\
    W(r, \theta)
    &=
    % \frac{ \chi_W \sqrt{2} }{ 4 \pi \epsilon_0 r^3 } ( 1 - 3 \cos^2\theta ),  
    % -\frac{ \chi_W  }{ \pi \epsilon_0 r^3 } \sqrt{ \frac{ 2 \pi }{ 5 } } Y_{2,0}(\theta),
    -\frac{ d^{(0)}_{\tilde{1} \rightarrow \tilde{0}} d^{(0)}_{\tilde{1} \rightarrow \tilde{2}} }{ \pi \epsilon_0 r^3 } \sqrt{ \frac{ 2 \pi }{ 5 } } Y_{2,0}(\theta).
\end{align}
\end{subequations} 
For efficient shielding, the two resonant thresholds should actually be slightly detuned, enough to suppress inelastic $|{\tilde{1},0; \tilde{1},0}\rangle \rightarrow |{\tilde{0},0; \tilde{2},0}\rangle$ transitions. 
The detuning $\Delta{E} = (E_{\tilde{1}, \tilde{1}} - E_{\tilde{2}, \tilde{0}})$, is found to be optimal typically around tens of megahertz.   
The dipole-dipole interactions are, by comparison, typically much smaller only get up to $\sim 100$ kHz outside the shielding barrier. This separation of energies allows us to expand Eq.~(\ref{eq:shielded_adiabat}) to first order in the inverse detuning, as:
\begin{align} \label{eq:effective_DCshielded_potential}
    V_{\rm eff}(r,\theta)
    &\approx 
    % \frac{ d^{\tilde{1} \rightarrow \tilde{1}} d^{\tilde{1} \rightarrow \tilde{1}} }{ 4 \pi \epsilon_0 }
    \frac{ C_{3, {\rm d.c.}} (1 - 3 \cos^2\theta) }{ r^3 }
    +
    % \frac{ ( d^{\tilde{1} \rightarrow \tilde{0}} d^{\tilde{1} \rightarrow \tilde{2}} )^2  }{ 8 \pi^2 \epsilon_0^2 \Delta{E} }
    \frac{ C_{6, {\rm d.c.}} ( 1 - 3 \cos^2\theta )^2 }{ r^6 },
    % \nonumber\\
    % &=
    % \frac{ \chi_a }{ 4 \pi \epsilon_0 }
    % \left( -4 \sqrt{ \frac{ \pi }{ 5 } } Y_{2,0}(\theta, \phi) \right)
    % \frac{ 1 }{ r^3 } \nonumber\\
    % &\quad 
    % +
    % \frac{ \chi_W^2  }{ 8 \pi^2 \epsilon_0^2 \Delta{E} }
    % \left(
    % \frac{ 8 \sqrt{ \pi } }{ 5 } Y_{0,0}(\theta, \phi)
    % +
    % \frac{ 16 }{ 7 } \sqrt{ \frac{ \pi }{ 5 } } Y_{2,0}(\theta, \phi)
    % +
    % \frac{ 48 \sqrt{ \pi } }{ 55 } Y_{4,0}(\theta, \phi)
    % \right)
    % \frac{ 1 }{ r^6 },
\end{align}
identifying an effective first-order dipole-dipole coefficient $C_{3, {\rm d.c.}} = { d^{(0)}_{\tilde{1} \rightarrow \tilde{1}} d^{(0)}_{\tilde{1} \rightarrow \tilde{1}} / (4 \pi \epsilon_0) }$ and repulsive van der Waals coefficient $C_{6, {\rm d.c.}} \approx { ( d^{(0)}_{\tilde{1} \rightarrow \tilde{0}} d^{(0)}_{\tilde{1} \rightarrow \tilde{2}} )^2  / ( 8 \pi^2 \epsilon_0^2 \Delta{E} ) }$. All constant energy offsets to $V_{\rm eff}(r,\theta)$ have been ignored in Eq.~(\ref{eq:effective_DCshielded_potential}).

The form of $V_{\rm eff}$ above now makes clear that a repulsive $r^{-6}$ dispersive barrier is now present as a result of the simultaneous $|\tilde{N}, m_{\tilde{N}} \rangle = | \tilde{1}, 0 \rangle \rightarrow | \tilde{0}, 0 \rangle$ and $| \tilde{1}, 0 \rangle \rightarrow | \tilde{2}, 0 \rangle$ exchange process between two molecules, reorienting them to prevent approach into the short range.     
It should be pointed out that the effective potential in Eq.~(\ref{eq:effective_DCshielded_potential}) vanishes identically at the dipolar magic angle $\theta^{\star} = \cos^{-1}1/\sqrt{3} \approx 54.7^{\circ}$.
This peculiarity is but a failure of the current truncated basis approximation, where proper accounting of other close-by pair molecular states would result in a repulsive barrier for all incident angles \cite{Mukherjee25_arxiv}. 
We opt to ignore further of such details here as we are primarily concerned with the interactions at long-range that affect differential scattering and collective gas thermalization.

\subsection{ Collisional shielding with a microwave field \label{sec:MWshielding} }

An alternative means to achieve collisional shielding is through the use of oscillating electric fields, instead of static ones. 
This time-dependent alternative relies on driving rotational transitions that are typically in the microwave band, and so is referred to as ``microwave shielding" \cite{Karman18_PRL, Lassabliere18_PRL}.  
Although the premise is similar, microwave shielding employs avoided crossings as are observed in the rotating frame of the microwave.  

For a single molecule immersed in a bath of microwave photons, the Hamiltonian is given by
\begin{align}
    {\cal H}
    &=
    {\cal H}_{\rm rot} 
    +
    {\cal H}_{\rm mw}
    +
    {\cal H}_{\rm a.c.} \\
    &=
    B_{\rm rot} \boldsymbol{N}^2
    +
    \sum_{\nu}
    \hbar \omega_{\nu}
    \left(
    a_{\nu}^{\dagger} a_{\nu} - n_{0, \nu}
    \right)
    -
    \sum_{\nu}
    \frac{ d {\rm E}_{\nu} }{ 2 \sqrt{ n_{0, \nu} } }
    \left[
    ( \hat{\boldsymbol{d}} \cdot \hat{\boldsymbol{\varepsilon}}^*_{\nu} ) a_{\nu}^{\dagger} 
    +
    ( \hat{\boldsymbol{d}} \cdot \hat{\boldsymbol{\varepsilon}}_{\nu} ) a_{\nu}
    \right], \nonumber
\end{align}
where $a_{\nu}$ ($a_{\nu}^{\dagger}$) is the lowering (raising) operator for microwave mode $\nu$ with frequency $\omega_{\nu}$, $n_{0, \nu}$ is a reference number of photons (think of a large near infinite photon bath), and $\boldsymbol{{\rm E}}_{\nu}(t) = {\rm E}_{\nu} \Re{ \hat{\boldsymbol{\varepsilon}}_{\nu} e^{i \omega_{\nu} t} }$ is the electric field with complex polarization $\hat{\boldsymbol{\varepsilon}}_{\nu}$.   
Although we utilize $\nu$ to index the microwave mode, we will only consider the case of a single mode with $\sigma^+$ polarization, so that the subscript will be implicit and dropped for the remainder of this chapter.  

To handle the single molecule system, we consider the joint molecule-photon basis $|{ N, m_N}\rangle|{ n }\rangle$, which are each the eigenstates for their respective uncoupled Hamiltonians:
\begin{subequations} \label{eq:single_molecule_eigenenergies}
\begin{align}
    H_{\rm rot} |{ N, m_N}\rangle
    &=
    B_{\rm rot} N ( N + 1 )
    |{ N, m_N}\rangle, \\
    H_{\rm mw}
    |{ n }\rangle
    &=
    \hbar 
    \omega n
    |{ n }\rangle,
\end{align}
\end{subequations}
defining $n$ as the differential number of photons with respect to the reference number in the photon background $n_{0}$. 
Note that when considering two (or more) molecules, there is only a single photon bath shared between all molecules and not a bath for each one. 
Circular microwave polarization gives
$\hat{\boldsymbol{\varepsilon}} = -(\hat{\boldsymbol{x}} + i \hat{\boldsymbol{y}}) / \sqrt{2}$.
The off-diagonal elements, given by the light-matter Hamiltonian ${\cal H}_{\rm a.c.}$ are, therefore, only nonzero when $N' = N \pm 1$ and $n' = n \pm 1$: 
\begin{align}
    \langle{ n' }|\langle{ N', m'_N}|
    {\cal H}_{\rm a.c.}
    |{ N, m_N}\rangle|{ n }\rangle 
    &=
    % \sum_{\nu=1}^2
    \frac{ d {\rm E}_{\rm a.c.} }{ 2 }
    \bigg[
    \bra{ N', m'_N }
    \hat{\boldsymbol{d}} \cdot \hat{\boldsymbol{\varepsilon}}^*
    \ket{ N, m_N }
    \delta_{n', n + 1} \\
    &\quad\quad\quad\quad
    +
    \bra{ N', m'_N }
    \hat{\boldsymbol{d}} \cdot \hat{\boldsymbol{\varepsilon}}
    \ket{ N, m_N }
    \delta_{n', n - 1} 
    \bigg] \nonumber\\
    &=
    -
    % \sum_{\nu=1}^2
    \frac{ d {\rm E}_{\rm a.c.} }{ 2 }
    \bigg[
    \bra{ N', m'_N }
    \frac{ x - i y }{ \sqrt{2} r }
    \ket{ N, m_N }
    \delta_{n', n + 1} \nonumber\\
    &\quad\quad\quad\quad
    \:\:
    +
    \bra{ N', m'_N }
    \frac{ x + i y }{ \sqrt{2} r }
    \ket{ N, m_N }
    \delta_{n', n - 1} 
    \bigg], \nonumber
\end{align}
where the matrix elements are computed as
\begin{align}
    \bra{ N', m'_N }
    \frac{ x \pm i y }{ \sqrt{ 2 } r }
    \ket{ N, m_N } 
    &=
    \mp \sqrt{ \frac{ 4 \pi }{ 3 } }
    \int Y^*_{N', m'_N}(\Omega) 
    % \cos\theta 
    % Y_{1, 0}(\Omega)
    Y_{1, \pm 1}(\Omega)
    Y_{N, m_N}(\Omega) d\Omega \nonumber\\
    =
    \mp (-1)^{m'_N}
    & \sqrt{ (2 N' + 1) (2 N + 1) } 
    \begin{pmatrix}
        N' & 1 & N \\
        -m'_N & \pm 1 & m_N 
    \end{pmatrix}
    \begin{pmatrix}
        N' & 1 & N \\
        0 & 0 & 0 
    \end{pmatrix}.
\end{align}

\begin{figure}[ht]
    \centering
    \begin{minipage}{\textwidth}
    \centering
    \includegraphics[width=0.75\columnwidth]{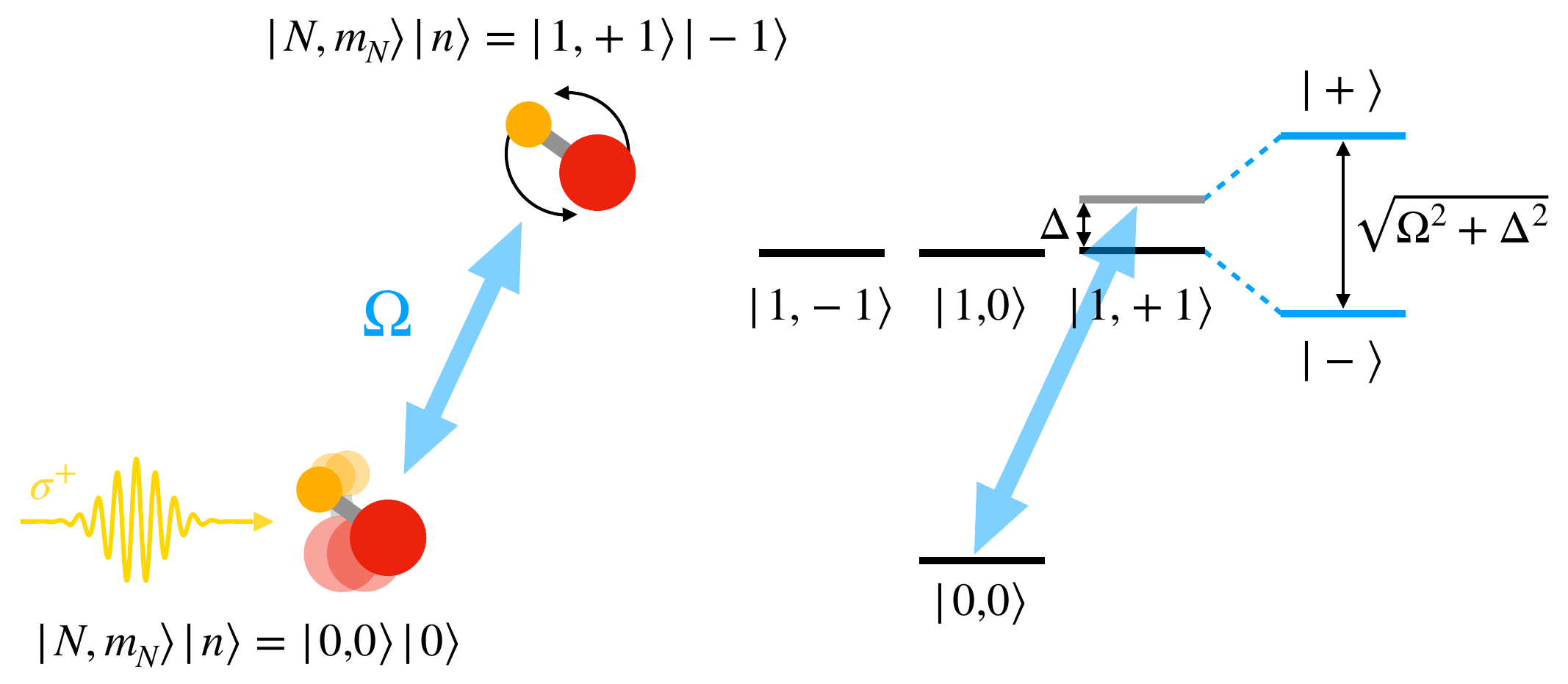}
    \caption{ Application of a coherent microwave drive on the $\ket{N, m_N} = \ket{ 0,0 }$ to $\ket{ 1,0 }$ molecular rotational transition with a blue-detuned, $\sigma^+$ polarized microwave. The right panel shows a schematic of the microwave dressed rotational levels, while the left panel provides an illustration of the process described in the main text.  
    % {\textcolor{red}{[Shouldn't bother me, but it does: both sets of blue arrow should slope from lower left to upper right.]}} 
    }
    \label{fig:microwave_dressing_cartoon}
    \end{minipage}
\end{figure}

The relevant basis states for a microwave at frequency close to the $\ket{ N, m_N } = \ket{ 0, 0 } \rightarrow \ket{ 1, +1 }$ transition can be taken as:
\begin{subequations}
\begin{align}
    \text{bright}: &\quad 
    \{ 
    |{ 0,0 }\rangle|{ 0 }\rangle, \:\:
    |{ 1,+1 }\rangle|{ -1 }\rangle
    \} \\
    \text{dark}: &\quad 
    \{ 
    |{ 1,0 }\rangle|{ -1 }\rangle, \:\:
    |{ 1,-1 }\rangle|{ -1 }\rangle
    \},
\end{align}
\end{subequations}
where bright and dark refer to states connected following the microwave polarization selection rules.
The bright states are Rabi-coupled, which might be intuited semiclassically as the perpetual absorbing then emitting of microwave photons by the molecule at a rate given by the Rabi frequency (illustrated in Fig.\ref{fig:microwave_dressing_cartoon}). Considering only these bright states, we obtain the reduced system  
\begin{align}
    \boldsymbol{{\cal H}}
    &\approx 
    \begin{pmatrix}
        0 & \hbar\Omega/2 \\ % & 0 \\ 
        \hbar\Omega/2 & -\hbar\Delta 
    \end{pmatrix},
\end{align}
having defined the Rabi frequency $\hbar\Omega = d {\rm E}_{\rm a.c.} / \sqrt{3}$,
microwave detuning $\Delta = \omega - \omega_{0 \rightarrow 1}$ with $\hbar\omega_{0 \rightarrow 1} = 2 B_{\rm rot}$, and
ignoring any constant energy offset. This two-state system is solved analytically by the spectrum: 
\begin{subequations} \label{eq:onebody_spectrum}
\begin{align}
    \ket{ - }
    &=
    \cos\varphi |{ 0, 0}\rangle|{ 0 }\rangle
    -
    \sin\varphi |{ 1, +1}\rangle|{ -1 }\rangle, 
    \quad 
    \varepsilon_{-}
    =
    -\frac{ \hbar\Delta }{ 2 }
    -
    \frac{ \hbar }{ 2 }
    \sqrt{ \Delta^2 + \Omega^2 }, \\
    \ket{ + }
    &=
    \sin\varphi |{ 0, 0}\rangle|{ 0 }\rangle
    +
    \cos\varphi |{ 1, +1}\rangle|{ -1 }\rangle,
    \quad 
    \varepsilon_{+}
    =
    -
    \frac{ \hbar\Delta  }{ 2 }
    +
    \frac{ \hbar }{ 2 }
    \sqrt{ \Delta^2 + \Omega^2 }, 
\end{align}
\end{subequations}
where we define the mixing angle:
\begin{align}
    \varphi
    &=
    \tan^{-1}
    \left(
    \frac{ \Delta }{ \Omega }
    +
    \sqrt{ 1 + \left( \frac{ \Delta }{ \Omega } \right)^2 }
    \right). 
\end{align}
At zero detuning, we get that the eigenstates of this system are equal antisymmetric and symmetric superpositions of the two bright basis states, while introducing detuning allows us to vary the relative amplitudes of each bare state.   
Similar to the case of d.c. shielding in Eq.~(\ref{eq:DC_dipole_moment}), an effective a.c. dipole moment relevant to first-order dipole-dipole interactions between microwave dressed molecules can be assigned:
\begin{align} \label{eq:effective_ac_dipole}
    d_{\rm a.c.}
    &=
    -\frac{ \partial \varepsilon_+ }{ \partial {\rm E}_{\rm a.c.} }
    =
    -\frac{ d }{ \sqrt{ 12 ( 1 + \Delta^2/\Omega^2 ) } },
\end{align}
assuming the microwave drives the $\ket{ N, m_N } = \ket{ 0, 0 } \rightarrow \ket{ 1, +1 }$ transition.

For the purposes of dipolar collisions, we should now utilize the joint two-molecule $+$ photon $+$ partial wave basis:
% \begin{align}
%     \ket{ N_A, m_{N_A}; N_B, m_{N_B} }\ket{ n }\ket{ L, m_L },
% \end{align}
$\ket{ N_A, m_{N_A}; N_B, m_{N_B} }\ket{ n }\ket{ L, m_L }$,
where $\ket{ L, m_L }$ are the collisional partial waves between molecules $A$ and $B$. 
In the collisions of identical molecules, these states must be appropriately symmetrized (antisymmetrized) under the exchange of identical bosons (fermions). We focus our discussions here to the symmetric two-molecule rotational sector:
\begin{align}
    &\ket{ N_A, m_{N_A}; N_B, m_{N_B} }_S\ket{ n }\ket{ L, m_L } \nonumber\\ 
    &\quad\quad\quad\quad
    =
    \frac{ \ket{ N_A, m_{N_A}; N_B, m_{N_B} } + \ket{ N_B, m_{N_B}; N_A, m_{N_A} } }{ \sqrt{ 2 (1 + \delta_{N_A, N_B} \delta_{m_{N_A}, m_{N_B}}) } }
    \ket{ n }\ket{ L, m_L },
\end{align}
with only even (odd) partial waves for identical bosons (fermions).

The close-to-resonant microwave dressing induces oscillating dipoles in the molecules, so that when they approach, they experience first-order dipole-dipole interactions. 
These dipole-dipole interactions occur between states with the same photon number and opposite parity in both molecular states: 
\begin{align}
    & \bra{ L', m'_L } 
    \langle n' |\langle N'_A, m'_{N_A}; N'_B, m'_{N_B} |
    V_{\rm dd}(\boldsymbol{r})
    | N_A, m_{N_A}; N_B, m_{N_B} \rangle| n \rangle 
    \ket{ L, m_L } \nonumber\\
    &=
    \bra{ L', m'_L } 
    \langle N'_A, m'_{N_A}; N'_B, m'_{N_B} |
    V_{\rm dd}(\boldsymbol{r})
    | N_A, m_{N_A}; N_B, m_{N_B} \rangle
    \ket{ L, m_L }
    \delta_{n', n},
\end{align}
with matrix elements in the bare rotational state basis provided in Eq.~(\ref{eq:DDI_matrix_elements}).

Before even considering the light-matter and matter-matter couplings, we can see that many of the rotor-rotor-photon eigenstates are far detuned by multiples of $B_{\rm rot}$ or $\hbar\omega$, from those that are directly coupled by the microwaves.  
Pruning away these energetically distant states amounts to a rotating wave approximation, which we make here. Notably, there are still states that are dark to the microwaves, but only separated from the bright states by energies on the order of the detuning and Rabi frequency. These states must also be included as they will be relevant when the two molecules interact.
We can further truncate our basis by noting that there are no dipole-dipole interactions among the doubly excited $\ket{ n = -2 }$ states. Moreover, the $\sigma^+$-polarized microwave field couples the ground state only to $m_N = +1$ excited states. Therefore, doubly excited states with both $m_{N_A}$ and $m_{N_B} \neq +1$ are completely uncoupled and can be removed from the basis. 
Finally, we are left with the basis set: 
\begin{align} \label{eq:2molecule_symmetrized_basis}
    \begin{tabular}{l|c}
        $\ket{ N_A, m_{N_A}; N_B, m_{N_B} }_S\ket{ n }$ & energy \\
        \hline
        $\ket{ 0,0; 0,0 }_S\ket{ 0 }$ & 0 \\
        $\ket{ 1,-1; 0,0 }_S\ket{ -1 }$ & $-\hbar\Delta$ \\
        $\ket{ 1,0; 0,0 }_S\ket{ -1 }$ & $-\hbar\Delta$ \\
        $\ket{ 1,+1; 0,0 }_S\ket{ -1 }$ & $-\hbar\Delta$ \\
        $\ket{ 1,+1; 1,-1 }_S\ket{ -2 }$ & $-2 \hbar\Delta$ \\
        $\ket{ 1,+1; 1,0 }_S\ket{ -2 }$ & $-2 \hbar\Delta$ \\
        $\ket{ 1,+1; 1,+1 }_S\ket{ -2 }$ & $-2 \hbar\Delta$ 
    \end{tabular},
\end{align}
in which the noninteracting dimer Hamiltonian ${\cal H}_{A,B} = {\cal H}_A + {\cal H}_B$ has the matrix representation
\begin{align} \label{eq:noninteracting_HAB}
    \boldsymbol{{\cal H}}_{A,B}(\boldsymbol{r})
    &=
    \begin{pmatrix}
        0 & 0 & 0 & \frac{\hbar\Omega}{\sqrt{ 2 }} & 0 & 0 & 0 \\
        0 & -\hbar\Delta & 0 & 0 & \frac{\hbar\Omega}{2} & 0 & 0 \\
        0 & 0 & -\hbar\Delta & 0 & 0 & \frac{\hbar\Omega}{2} & 0 \\
        \frac{\hbar\Omega}{\sqrt{ 2 }} & 0 & 0 & -\hbar\Delta & 0 & 0 & \frac{\hbar\Omega}{\sqrt{ 2 }} \\
        0 & \frac{\hbar\Omega}{2} & 0 & 0 & -2\hbar\Delta & 0 & 0 \\
        0 & 0 & \frac{\hbar\Omega}{2} & 0 & 0 & -2\hbar\Delta & 0 \\
        0 & 0 & 0 & \frac{\hbar\Omega}{\sqrt{ 2 }} & 0 & 0 & -2\hbar\Delta
    \end{pmatrix}.
\end{align}
Diagonalizing the matrix above gives the energetic thresholds at large $r$, that follow directly as linear combinations of the one-molecule microwave dressed eigenenergies (\ref{eq:onebody_spectrum}):
\begin{align} \label{eq:2molecule_ACsymmetrized_basis}
    \begin{tabular}{l|c}
        dressed state & energy \\
        \hline
        $\ket{ -; - }_S$ & 
        $2 \varepsilon_{-}$ \\
        $\ket{ -; 1,-1 }_S$ & 
        $\varepsilon_{-} - \hbar\Delta$ \\
        $\ket{ -; 1,0 }_S$ & $\varepsilon_{-} - \hbar\Delta$ \\
        $\ket{ -; + }_S$ &  $\varepsilon_{-} + \varepsilon_{+}$ \\
        $\ket{ +; 1,-1 }_S$ & $\varepsilon_{+} - \hbar\Delta$ \\
        $\ket{ +; 1,0 }_S$ & $\varepsilon_{+} - \hbar\Delta$ \\
        $\ket{ +; + }_S$ & $2 \varepsilon_{+}$ 
    \end{tabular},
\end{align}
where $\varepsilon_{\pm} = -\hbar\Delta/2 \pm (\hbar/2)\sqrt{ \Delta^2 + \Omega^2 }$, identifying five unique threshold energies.

\begin{figure}[ht]
    \centering
    \begin{minipage}{\textwidth}
    \centering
    \includegraphics[width=\columnwidth]{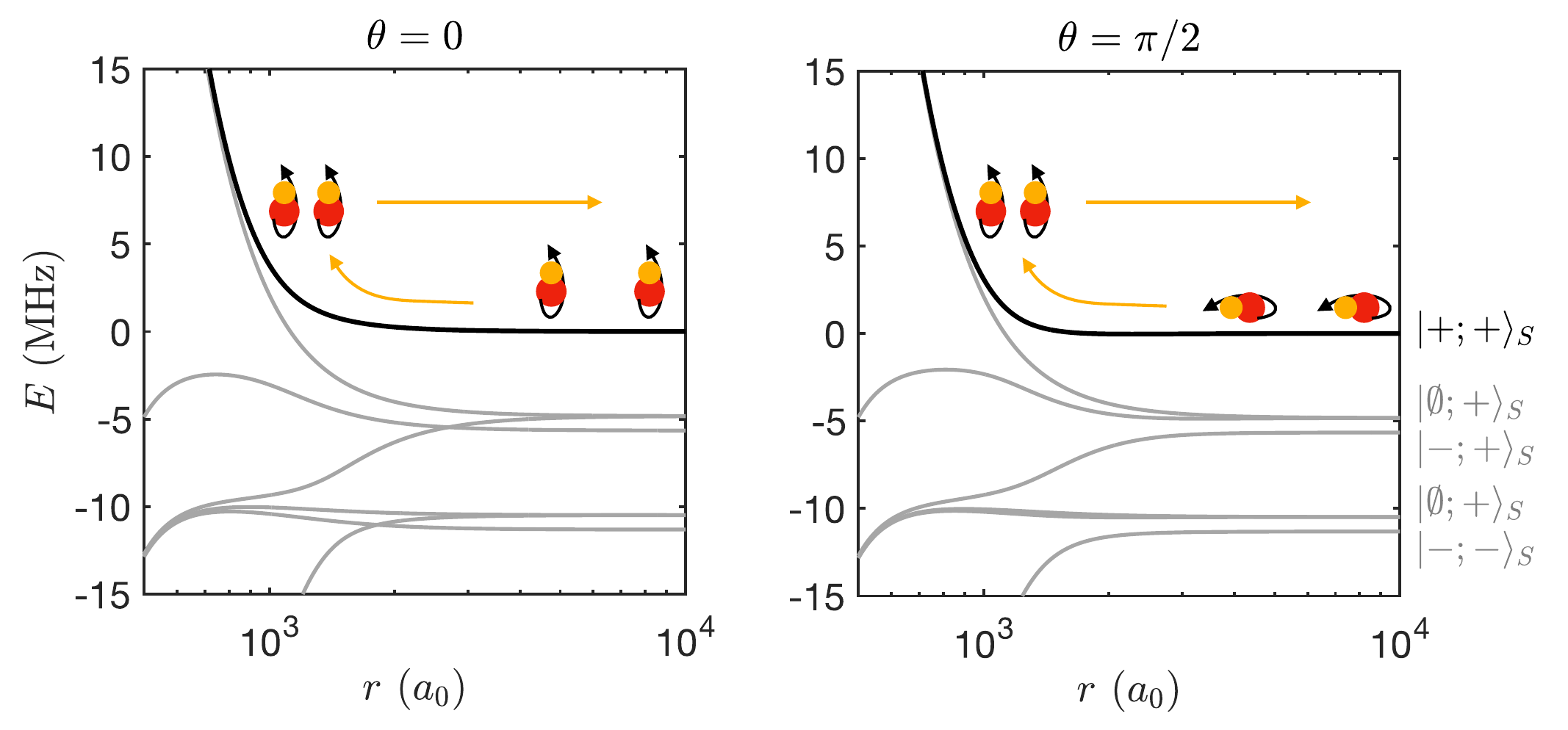}
    \caption{ Microwave dressed potentials (solid curves) plotted in real space between NaCs molecules, obtained with a $\Delta = 2\pi \times 4$ MHz detuning and a Rabi frequency of $\Omega = 2 \pi \times 4$ MHz. 
    The highest potential, responsible for microwave shielding, is plotted and labeled in black. All other adiabats are plotted and labeled in gray, with $\ket{ \emptyset } = \ket{ 1, -1 }, \ket{ 1, 0 }$ denoting states that are dark to the microwave.   
    The angle $\theta$ denotes the incident scattering angle, as represented by the schematic orientations of the illustrated rotating molecules and their trajectories. }
    \label{fig:adiabatsNaCs_Omega4MHz_Delta4MHz}
    \end{minipage}
\end{figure}

As interactions become significant at smaller $r$, the interacting adiabatic Hamiltonian $H_{\rm ad}(\boldsymbol{r}) = {\cal H}_A + {\cal H}_B + V_{\rm dd}(\boldsymbol{r})$, then requires inclusion of the dipole-dipole couplings (\ref{eq:DDI_matrix_elements}), which is given in the basis of Eq.~(\ref{eq:2molecule_symmetrized_basis}) as: 
\begin{align} \label{eq:DDI_coupling}
    \boldsymbol{V}_{\rm dd}(\boldsymbol{r})
    &=
    \begin{pmatrix}
        0 & 0 & 0 & 0& 0\quad & 0\quad & 0 \\
        0 & \frac{1}{3} \frac{ d^2 C_{2,0}(\theta, \phi) }{ 4 \pi \epsilon_0 r^3 } & -\frac{1}{\sqrt{3}} \frac{ d^2 C_{2,+1}(\theta, \phi) }{ 4 \pi \epsilon_0 r^3 } & \sqrt{ \frac{2}{3} } \frac{ d^2 C_{2,+2}(\theta, \phi) }{ 4 \pi \epsilon_0 r^3 } & 0\quad & 0\quad & 0 \\
        0 & \frac{1}{\sqrt{3}} \frac{ d^2 C_{2,-1}(\theta, \phi) }{ 4 \pi \epsilon_0 r^3 } &  -\frac{2}{3} \frac{ d^2 C_{2,0}(\theta, \phi) }{ 4 \pi \epsilon_0 r^3 } & \frac{1}{\sqrt{3}} \frac{ d^2 C_{2,+1}(\theta, \phi) }{ 4 \pi \epsilon_0 r^3 } & 0\quad & 0\quad & 0 \\
        0 & \sqrt{ \frac{2}{3} } \frac{ d^2 C_{2,-2}(\theta, \phi) }{ 4 \pi \epsilon_0 r^3 } & -\frac{1}{\sqrt{3}} \frac{ d^2 C_{2,-1}(\theta, \phi) }{ 4 \pi \epsilon_0 r^3 } & \frac{1}{3} \frac{ d^2 C_{2,0}(\theta, \phi) }{ 4 \pi \epsilon_0 r^3 } & 0\quad & 0\quad & 0 \\
        0 & 0 & 0 & 0 & 0\quad & 0\quad & 0 \\
        0 & 0 & 0 & 0 & 0\quad & 0\quad & 0 \\
        0 & 0 & 0 & 0 & 0\quad & 0\quad & 0
    \end{pmatrix}. 
\end{align}
Diagonalizing the adiabatic Hamiltonian of two microwave-dressed interacting NaCs molecules gives the potential energy curves in Fig.~\ref{fig:adiabatsNaCs_Omega4MHz_Delta4MHz}, with Rabi frequency and detuning $\Omega = \Delta = 2\pi \times 4$ MHz. 
We see that the uppermost adiabat (solid black curve) is repulsive in both orientations presented, resulting in collisional shielding.

\begin{figure}[ht]
    \centering
    \begin{minipage}{\textwidth}
    \centering
    \includegraphics[width=0.8\linewidth]{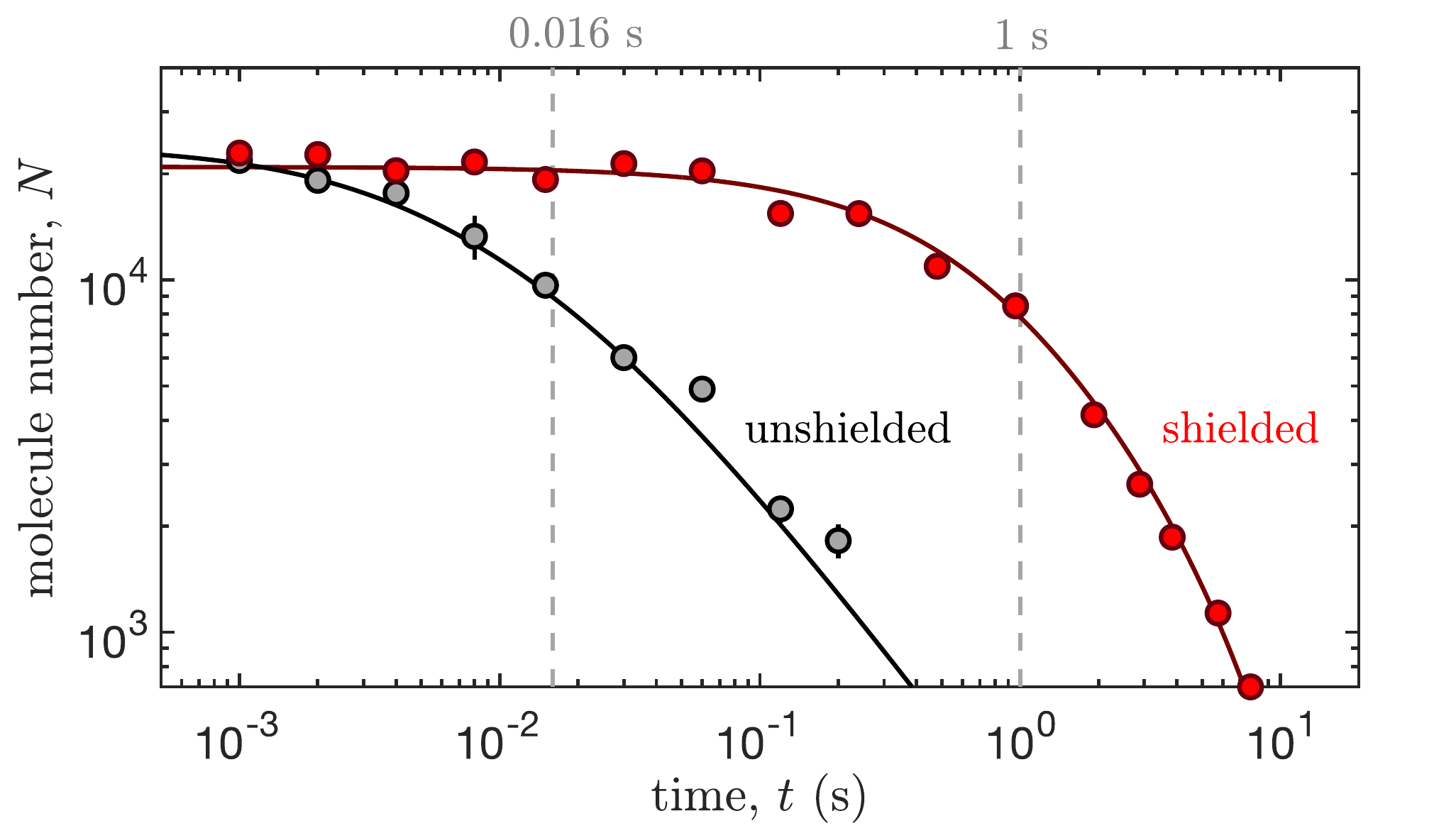}
    \caption{ Molecule number loss vs time for an ultracold gas of NaCs molecules with (red) and without (black) microwave shielding. The gray dashed lines indicate the respective $1/e$ lifetimes of the molecular samples. Error bars indicate the 1$\sigma$ standard-error-of-the-mean from ten measurement repetitions. Microwave shielding is performed with $\Omega/(2\pi) = 4$ MHz and $\Delta/\Omega = 1$.
    This plot is reproduced with data from Ref.~\cite{Bigagli23_NatPhys}. }
    \label{fig:Bigagli23_NatPhys_Fig1a}
    \end{minipage}
\end{figure}

Despite the high shielding barriers, inelastic transitions to other nearby rotational states or tunneling through the shielding barrier into the short-range can still occur during molecular collisions. 
These processes result in a finite efficacy of microwave shielding, 
resulting in a non-negligible but small amount of two-body loss.  
Nevertheless, microwave shielding has allowed the production of collisionally stable ultracold samples of polar molecules with second-scale lifetimes, exhibited by the number loss curves in Fig.~\ref{fig:Bigagli23_NatPhys_Fig1a}, with data obtained from Ref.~\cite{Bigagli23_NatPhys}.  
In the figure, we see that the presence of monochromatic microwave radiation extends the lifetime of a NaCs molecular cloud by around two orders of magnitude, similarly observed in gases of bosonic NaRb \cite{Lin23_PRX} and fermionic NaK \cite{Schindewolf22_Nat}.

\subsubsection{ Effective microwave shielded potential }

With the remarkable suppression of loss achievable with microwave shielding, attention can be drawn to elastic scattering between molecules.  
As done for static field shielding, it is useful to derive a single-channel effective potential by considering only the lowest-order perturbation terms obtained from the interaction potential in Eq.~(\ref{eq:DDI_coupling}). 
At long-range, the first-order shielded dipole-dipole interaction is obtained from the upper most eigenstate $\ket{ +; + }_S$, that follows from Eq.~(\ref{eq:noninteracting_HAB}) as:
\begin{align}
    \ket{ +; + }_S
    &=
    \frac{ 1 }{ 
    2 \sqrt{ \left( 1 + \delta^2 \right) 
    \big[
    1
    +
    2 \delta 
    \big( 
    \delta + \sqrt{ 1 + \delta^2 } 
    \big)
    \big] }
    }
    \begin{pmatrix}
        1 + 2 \delta \left( \delta + \sqrt{ 1 + \delta^2 } \right) \\
        0 \\
        0 \\
        \sqrt{2} \left( \delta + \sqrt{ 1 + \delta^2 } \right) \\
        0 \\
        0 \\
        1 
    \end{pmatrix}, 
\end{align}
where $\delta = \Delta/\Omega$, 
granting us the simple expression
\begin{align} \label{eq:microwave_effectiveDDI}
    V^{(1)}_{\rm dd}(\boldsymbol{r})
    &=
    _S\bra{ +; + }
    {V}_{\rm dd}(\boldsymbol{r})
    \ket{ +; + }_S
    =
    \frac{ 1 }{ 12 ( 1 + \delta^2 ) }
    \frac{ d^2 }{ 4 \pi \epsilon_0 }
    \frac{ 3 \cos^2\theta - 1 }{ r^3 }.
\end{align}
Most notably, the form of Eq.~(\ref{eq:microwave_effectiveDDI}) is very similar to Eq.~(\ref{eq:point_DDI}) although differing by an overall minus sign, the so-called \textit{anti-dipolar} interactions. 
This is most intuitively understood with a semiclassical picture of rotating dipoles.  In this case of circularly polarized microwaves, we can consider the dipoles as pointing orthogonal to the $z$ axis and rotating in the $x$-$y$ plane at the frequency of the microwave, thus: ${\bf d}_{1,2} = d( \cos \omega t {\hat x} + \sin \omega t {\hat y})$.  Then dipoles separated by a radius vector ${\bf r} = r(\sin \theta \cos \phi, \sin \theta \sin \phi , \cos \theta)$ would experience a dipole-dipole interaction (\ref{eq:DDI_potential}):
\begin{align}
    V_{\rm dd}(\boldsymbol{r}, t) 
    &=
    \frac{ d^2 }{ 4 \pi \epsilon_0 r^3 } 
    \Big( 1 - 3 \sin^2 \theta \cos^2 ( \omega t - \phi) \Big).
\end{align}
For rapid rotation, we can average the squared cosine over a period of rotation, yielding a factor $1/2$, whereby the angular dependence of the interaction is $1-(3/2)\sin^2 \theta = (1/2)(3 \cos^2 \theta - 1)$.  
This is half the size and of the opposite sign of the familiar interaction between two dipoles statically polarized along $z$.

Similarly, we can obtain a simple expression for the second-order interaction by taking only the three energetically closest states to $\ket{ +; + }_S$, resulting in: 
\begin{align}
    V^{(2)}_{\rm dd}(\boldsymbol{r})
    &\approx 
    \frac{ | _S\bra{ +; 1, -1 } {V}_{\rm dd}(\boldsymbol{r}) \ket{ +; + }_S |^2 }{ 2\varepsilon_{+} - ( \varepsilon_{+} - \hbar\Delta ) }
    +
    \frac{ | _S\bra{ +; 1, 0 } {V}_{\rm dd}(\boldsymbol{r}) \ket{ +; + }_S |^2 }{ 2\varepsilon_{+} - ( \varepsilon_{+} - \hbar\Delta ) } \nonumber\\
    &\quad
    +
    \frac{ | _S\bra{ +; - } {V}_{\rm dd}(\boldsymbol{r}) \ket{ +; + }_S |^2 }{ 2\varepsilon_{+} - ( \varepsilon_{+} + \varepsilon_{-} ) } \nonumber\\
    &=
    \left( \frac{ d^2 }{ 4 \pi \epsilon_0 } \right)^2 
    \Bigg(
    \frac{ 1 }{ 8 \hbar\Omega ( 1 + \delta^2 )^{3/2} }
    \frac{ ( 1 + \cos^2\theta ) \sin^2\theta }{ r^6 } \nonumber\\
    &\quad\quad\quad\quad
    \quad\quad
    +
    \frac{ \delta^2 }{ 72 \hbar\Omega  \left( 1 + \delta^2 \right)^{5/2} }
    \frac{  \left( 1 - 3 \cos^2\theta \right)^2}{ r^6 }
    \Bigg).
\end{align}
Putting the two expansion terms together, we obtain the effective microwave shielded potential:
\begin{align} \label{eq:effective_ACshielded_potential}
    V_{\rm eff}(\boldsymbol{r})
    &=
    % \frac{ d_{\rm a.c.}^2 }{ 4 \pi \epsilon_0 }
    \frac{ C_{3, {\rm a.c.}} ( 3 \cos^2\theta - 1 ) }{ r^3 } \nonumber\\
    &\quad 
    +
    \frac{ C_{6, {\rm a.c.}} }{ r^6 }
    \left[
    ( 1 + \cos^2\theta ) \sin^2\theta
    +
    \frac{ \delta^2 }{ 9 ( 1 + \delta^2 ) }
    ( 1 - 3\cos^2\theta )^2
    \right],
\end{align}
where $C_{3, {\rm a.c.}} = d_{\rm a.c.}^2 / (4 \pi \epsilon_0)$ is the first-order dipole-dipole coefficient with effective a.c. dipole moment $d_{\rm a.c.} = d / \sqrt{ 12 ( 1 + \delta^2 ) }$, and $C_{6, {\rm a.c.}} = [ d^2/(4 \pi \epsilon_0) ]^2 [ 8 \hbar\Omega ( 1 + \delta^2 )^{3/2} ]^{-1}$ is the effective dispersion interaction.
Ref.~\cite{Deng23_PRL} derives a similar microwave shielded effective potential, although only the first dispersion term above is kept there.

Pertaining to the rethermalization of a microwave shielded molecular gas, the flipped sign on the leading order dipole-dipole interaction leaves the number of collisions per rethermalization unchanged from Eqs.~(\ref{eq:NCPR_fermions}), since the differential cross section involves a square of the scattering amplitude (\ref{eq:DCS_scatAmp2}). 
Albeit negligible for identical fermions, the shorter ranged dispersive interactions can significantly affect the scattering of shielded identical bosonic molecules--the result of a barrier-less $s$-wave channel. In particular, the $s$-wave scattering length of molecular bosons can be tuned with the strength of the microwaves \cite{Lassabliere18_PRL, Quemener23_PRL}, a consequence of changing both the short-range shielding barrier and long-range dipolar tail.

\subsection{ Implications of shielded interactions to relaxation }

\begin{figure}[ht]
    \centering
    \begin{minipage}{\textwidth}
    \centering
    \includegraphics[width=\linewidth]{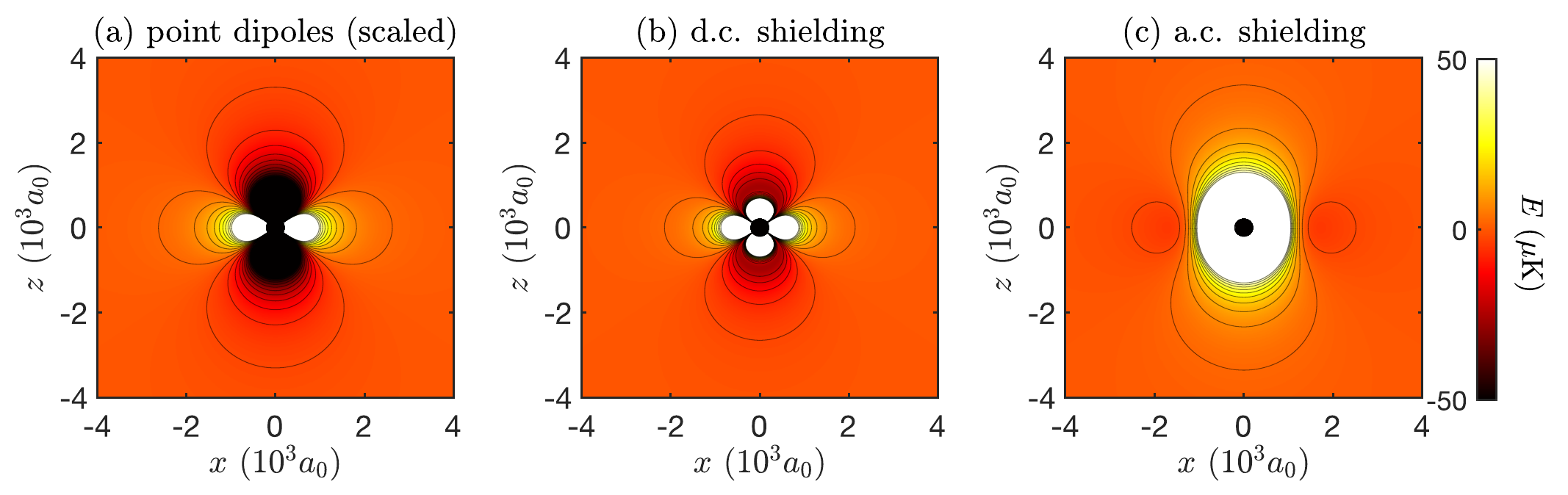}
    \caption{ Comparison of the potential energy surfaces along $y=0$ for (a) point-dipoles, (b) d.c. Forster resonant shielded polar molecules, and (c) a.c. field shielded polar molecules. The point-dipole potential has the dipole moment scaled by 0.2, for ease of comparison. The color map saturates at $\pm 50$ $\mu$K, and anything within $r \leq 200 a_0$ (approximately the electronic van der Waals length) is colored black to denote short-range absorption. 
    % \textcolor{red}{[this is glorious!]}
    }
    \label{fig:effective_potentials}
    \end{minipage}
\end{figure}

The derived effective shielded potentials of the preceding sections allow their elastic interactions to be placed on equal footing, illustrated in Fig.~\ref{fig:effective_potentials}. Adopting the example of NaCs molecules--with molecular frame dipole moment $d = 4.7$ D--the figure compares the potential energy surfaces for (a) point-dipoles, (b) static field F\"orster resonant shielded molecules, and (c) circularly polarized microwave shielded molecules in the respective subplots, a striking showcase of their differing strengths and angular character. 
In subplot (b), static field shielding requires the molecules to be prepared in the first rotationally excited state, possessing a relatively small induced dipole moment of $d_{\tilde{1}\rightarrow\tilde{1}}^{(0)} \approx -0.68$ D at the shielding field ${\rm E}_{\rm d.c.} = 2.4$ kV/cm. 
Meanwhile, microwave shielding also effectively suppresses the molecular frame dipole moment to $d_{\rm a.c.} \approx 0.96$ D, in the rotating frame of the microwaves with $\Delta/\Omega = 1$ (\ref{eq:effective_ac_dipole}).
For the case of point-dipoles, we artificially scaled down the molecular frame dipole moment by a factor of 0.2 for ease of visual comparison.

\begin{figure}[ht]
    \centering
    \begin{minipage}{\textwidth}
    \centering
    \includegraphics[width=0.8\linewidth]{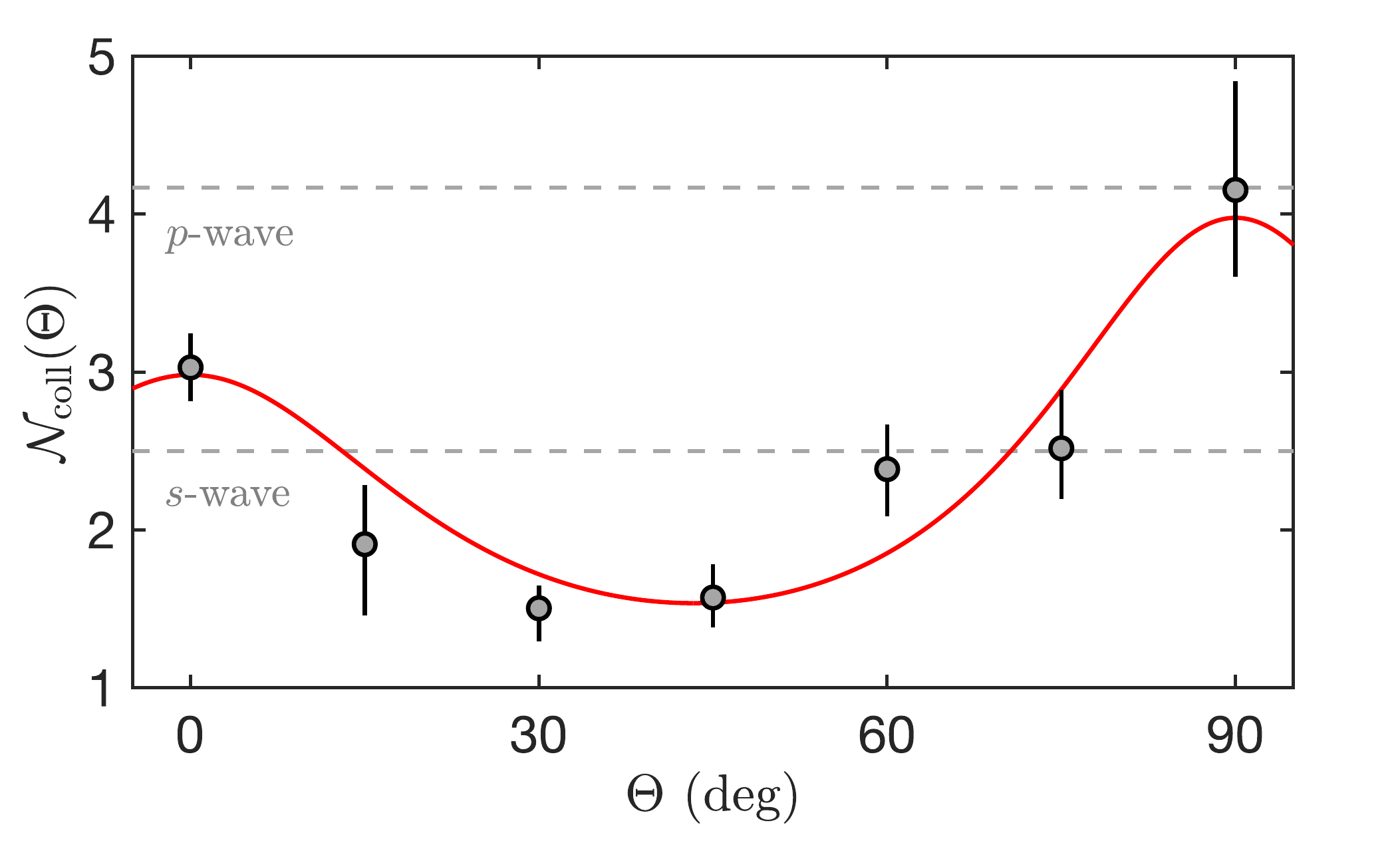}
    \caption{ A comparison of the number of collisions per rethermalization between the theoretical prediction (solid red curve) and experimental data obtained from the JILA KRb cross-dimensional rethermalization experiment \cite{Li21_NatPhys}.  
    The excitation and thermalization axes here are coincident, along the axis defined by $\Theta = 0$. We also plot the number of collisions per rethermalization for $s$- and $p$-wave colliders as dashed horizontal lines, labeled respectively. This plot is reproduced with data from Ref.~\cite{Wang21_PRA}.  }
    \label{fig:Ncoll_KRb2021}
    \end{minipage}
\end{figure}

Despite the differences observed in Fig.~\ref{fig:effective_potentials}, equations (\ref{eq:effective_DCshielded_potential}) and (\ref{eq:effective_ACshielded_potential}) indicate that the interactions of shielded molecules at long-range are still just those between point-dipoles, up to a sign for microwave shielding. 
As a result, scattering observables of shielded polar molecules in the low energy threshold regime are practically identical to those between point-dipoles, 
rendering the results obtained in Sec.~(\ref{sec:DDI_rethermalization}) valid for bulk gases of ultracold shielded molecules.   
In fact, the analytic formulas for ${\cal N}$ (\ref{eq:NCPR_fermions}) have stood up against comparisons with experimental rethermalization measurements with static field shielded fermionic $^{40}$K$^{87}$Rb molecules at JILA. In that experiment, detailed in Ref.~\cite{Li21_NatPhys}, the relevant number of collisions per rethermalization was: 
\begin{align}
    N_{\rm coll}(\Theta)
    &=
    \frac{ 112 \lambda }{ 45 + 4 \cos(2\Theta) - 17 \cos(4\Theta) },
\end{align}
where $\lambda \approx 0.87$ is an additional factor accounting for a large initial temperature anisotropy \cite{Goldwin05_PRA}.  
Excellent agreement is found in a comparison of this theoretical formula (solid red curve) with the experimental data (points with error bars) provided in Fig.~\ref{fig:Ncoll_KRb2021}, bolstering the validity of Eqs.~(\ref{eq:NCPR_fermions}) and the point dipole approximation.

There are, however, circumstances where field-dressed long-range molecular interactions do result in observable deviations from the case of point-dipoles. 
One such instance is when microwave shielding is employed with microwaves with a small degree of ellipticity.   
It was found in Ref.~\cite{Karman19_PRA} that so long as the polarization is around $90\%$ circularly, microwaves can still be effective to prevent short-range molecular loss--great news given producing truly perfectly circular microwaves in the lab is very difficult in practice.    
But a nonzero ellipticity would introduce additional anisotropy to the long-range molecular interaction potential, potentially changing the number of collisions per rethermalization.   
It turns out that following a similar derivation to that in Sec.~\ref{sec:MWshielding}, the effective dipole-dipole interaction with elliptically polarized microwaves has the form \cite{Deng23_PRL}:
\begin{align}
    V_{\rm eff, dd}(\boldsymbol{r}; \xi)
    &=
    \frac{ C_{3, {\rm a.c.}} }{ r^3 }
    \left[
    3 \cos^2\theta - 1 + 3 \sin^2\theta \cos(2\phi) \sin(2\xi)
    \right],
\end{align}
valid for small ellipticity angles $\xi \leq 20^{\circ}$. 
Once again employing the Born approximation, we can derive the unsymmetrized close-to-threshold scattering amplitude for this interaction as:
\begin{align} \label{eq:ellipticalMW_scatAmp}
    f_{\rm eff}( \hat{\boldsymbol{k}}', \hat{\boldsymbol{k}} )
    &\approx
    -a_s
    -
    \frac{ 2 \mu }{ \hbar^2 }
    \frac{ 1 }{ 4 \pi }
    \int d^3 \boldsymbol{r}
    e^{ -i \boldsymbol{k}' \cdot \boldsymbol{r} }
    V_{\rm eff, dd}(\boldsymbol{r}; \xi)
    e^{ i \boldsymbol{k} \cdot \boldsymbol{r} } \nonumber\\
    % &=
    % -a_s
    % -
    % \frac{ 2 \mu }{ \hbar^2 }
    % \frac{ C_{3, {\rm a.c.}} }{ 4 \pi }
    % \int d^3 \boldsymbol{r}
    % \left(
    % 4 \sqrt{ \frac{ \pi }{ 5 } } Y_{2,0}(\hat{\boldsymbol{r}})
    % +
    % 6 \sqrt{ \frac{ 2\pi }{ 15 } } Y_{2,\pm 2}(\hat{\boldsymbol{r}})
    % \sin( 2\xi )
    % \right)
    % e^{ i (\boldsymbol{k} - \boldsymbol{k}') \cdot \boldsymbol{r} } \nonumber\\
    % &=
    % -a_s
    % -
    % \frac{ 2 \mu C_{3, {\rm a.c.}} }{ \hbar^2 }
    % \sum_{ \ell, m_{\ell} }
    % i^{\ell} Y_{\ell, m_{\ell}}^*(\hat{\boldsymbol{q}})
    % \int d^3 \boldsymbol{r}
    % \left(
    % 4 \sqrt{ \frac{ \pi }{ 5 } } Y_{2,0}(\hat{\boldsymbol{r}})
    % +
    % 6 \sqrt{ \frac{ 2\pi }{ 15 } } Y_{2,\pm 2}(\hat{\boldsymbol{r}})
    % \sin( 2\xi )
    % \right)
    % j_{\ell}(q r) 
    % Y_{\ell, m_{\ell}}(\hat{\boldsymbol{r}}) \nonumber\\
    % &=
    % -a_s
    % -
    % \frac{ 2 \mu C_{3, {\rm a.c.}} }{ \hbar^2 }
    % \int d r \frac{ j_{2}(q r) }{ r }
    % \left(
    % 4 \sqrt{ \frac{ \pi }{ 5 } } 
    % Y_{2,0}^*(\hat{\boldsymbol{q}})
    % +
    % 6 \sqrt{ \frac{ 2\pi }{ 15 } } 
    % Y_{2,\pm 2}(\hat{\boldsymbol{q}})
    % \sin( 2\xi )
    % \right) \nonumber\\
    % &=
    % -a_s
    % -
    % \frac{ 2 \mu C_{3, {\rm a.c.}} }{ \hbar^2 }
    % \left(
    % \frac{ 4 }{ 3 } \sqrt{ \frac{ \pi }{ 5 } } 
    % Y_{2,0}^*(\hat{\boldsymbol{k}} - \hat{\boldsymbol{k}}')
    % +
    % 2 \sqrt{ \frac{ 2\pi }{ 15 } } 
    % Y_{2,\pm 2}(\hat{\boldsymbol{k}} - \hat{\boldsymbol{k}}')
    % \sin( 2\xi )
    % \right) \nonumber\\
    &=
    -a_s
    -
    \frac{ \mu C_{3, {\rm a.c.}} }{ \hbar^2 }
    \Bigg(
    - 
    \frac{ 2 }{ 3 }
    +
    [ 1 + \sin(2\xi) ]
    \frac{ ( \hat{\boldsymbol{k}}\cdot\hat{\boldsymbol{S}}_{\rm a.c.} - \hat{\boldsymbol{k}}'\cdot\hat{\boldsymbol{S}}_{\rm a.c.} )^2 }{ 1 - \hat{\boldsymbol{k}}\cdot\hat{\boldsymbol{k}}' } \nonumber\\
    &\quad\quad\quad\quad
    \quad\quad\quad\quad
    \quad
    +
    2 \sin(2\xi)
    \left[
    \frac{ ( \hat{\boldsymbol{k}}\cdot\hat{\boldsymbol{S}}_{\rm a.c.}^{\perp} - \hat{\boldsymbol{k}}'\cdot\hat{\boldsymbol{S}}_{\rm a.c.}^{\perp} )^2 }{ 1 - \hat{\boldsymbol{k}}\cdot\hat{\boldsymbol{k}}' }
    -
    1
    \right]
    \Bigg), 
\end{align}
where $\hat{\boldsymbol{S}}_{\rm a.c.}$ is the direction of microwave propagation and $\hat{\boldsymbol{S}}_{\rm a.c.}^{\perp}$ is the direction orthogonal to it, both assumed lying in the laboratory $x,z$ plane. 
By imposing the appropriate particle exchange symmetries, we obtain the ellipticity-modified integral cross sections:
\begin{subequations}
\begin{align}
    \overline{\sigma}_F 
    &=
    \frac{ 16 \pi a_d^2 }{ 15 }
    \left[ 5 - 3 \cos(4\xi) \right], \\
    \overline{\sigma}_B 
    &=
    8 \pi a_s^2
    +
    \frac{ 32\pi a_d^2 }{ 45 }
    \left[
    7 + 4\sin(2\xi) - 6\cos(4\xi)
    \right]
    -
    \frac{ 32\pi a_s a_d }{ 3 }
    \sin(2\xi).
\end{align}
\end{subequations}
Unlike the case of ordinary dipoles, bosonic molecules shielded by elliptically polarized microwaves can now have their $s$-wave scattering and dipole lengths interfere at the level of their integral cross sections.

The additional anisotropy resultant from nonzero ellipticity in the microwaves
thus inadvertently changes the nature of momentum transfer during molecular collisions.      
The result is that for microwaves propagating along $\hat{\boldsymbol{S}}_{\rm a.c.} = \hat{\boldsymbol{z}}$, the number of collisions per rethermalization ${\cal N}_{ij}$ can also be tuned by the microwave ellipticity $\xi$:
\begin{subequations}
\begin{align}
    {\cal N}_{xx}(\xi)
    &=
    \frac{ 14 [ 5 - 3\cos(4\xi) ] }{ 17 - 11\cos(4\xi) - 4\sin(2\xi) }, \\
    {\cal N}_{yy}(\xi)
    &=
    \frac{ 70 - 42\cos(4\xi) }{ 17 - 11\cos(4\xi) + 4\sin(2\xi) }, \\
    {\cal N}_{zz}(\xi)
    &=
    \frac{ 7 }{ 4 }
    \left(
    3
    +
    \frac{ 1 }{ \cos(4\xi) - 2 }
    \right), \\
    {\cal N}_{xy}(\xi)
    &=
    3
    +
    \frac{ 8 }{ 9 - 7\cos(4\xi) }, \\
    {\cal N}_{xz}(\xi)
    &=
    \frac{ 7 [ 5 - 3\cos(4\xi) ] }{ 4 [2 - \cos(4\xi) - \sin(2\xi)] }, \\
    {\cal N}_{yz}(\xi)
    &=
    \frac{ 7 [ 5 - 3\cos(4\xi) ] }{ 4 [2 - \cos(4\xi) + \sin(2\xi)] },
\end{align}
\end{subequations}
where the expressions above are provided for identical fermionic molecules. 

\begin{figure}[ht]
    \centering
    \begin{minipage}{\textwidth}
    \centering
    \includegraphics[width=\linewidth]{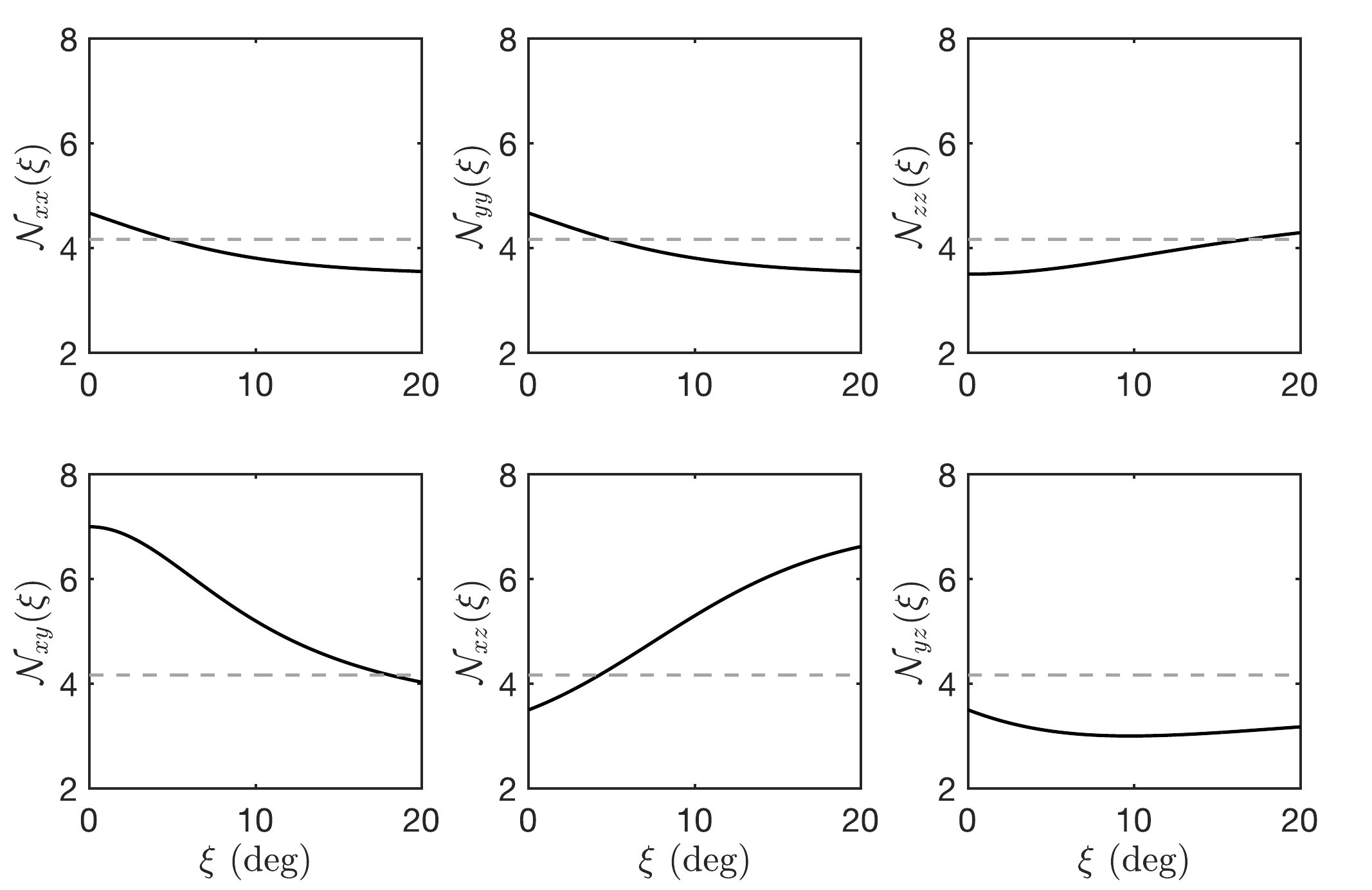}
    \caption{ The number of collisions per rethermalization for microwave shielded molecular fermions ${\cal N}_{i j}$ (solid black curves) as a function of microwave ellipticity $\xi$. For comparison, we also plot ${\cal N}_p = 25/6$ for $p$-wave scattering (dashed gray lines). }
    \label{fig:Ncoll_ellipticalMW}
    \end{minipage}
\end{figure}

Results for the number of collisions per rethermalization are shown in Fig.~\ref{fig:Ncoll_ellipticalMW}, as functions of the ellipticity. In all cases the ellipticity remains small, meaning that we are considering the limit where shielding remains effective.  Also shown for comparison is the value for $p$-wave scattering (dashed line).  These results verify that there are cases where ellipticity actually enhances the rate of thermalization.

\section{ Discussion and remarks }

The modern world of experimentally-controlled ultracold molecular gases is replete with opportunities to control the form of two-body interactions, and thus the many-body behavior of the gas. Manipulation by d.c. electric fields, or by circularly- or elliptically-polarized microwaves, bear similarities, yet each has its own idiosyncrasies, which we have here emphasized by presenting them together on a common footing.  Paramount among this control is the ability to thwart chemical reactions, thus preserving the mechanical integrity of the gas. Within the need to do so, there yet remain opportunities to direct elastic scattering to influence anisotropic rethermalization dynamics, for purposes of collisional measurements, nonequilibrium dynamics, and efficient evaporative cooling strategies.  The results and formulas provided herein are intended to help guide such efforts.

The subject can be taken further in the future.  
Ignored in this chapter is the ability to apply multiple microwaves simultaneously at different frequencies and polarizations \cite{Karman25_PRXQ}, or combinations of both static and microwave electric fields \cite{Gorshkov08_PRL, Karman18_PRL}.  The former has proven necessary in some experiments to prevent loss due to three-body recombination \cite{Bigagli24_Nat, Shi25_arxiv}, and has been evaluated as a novel form of two-photon-assisted collisions \cite{Stevenson25_inprep}. 
Strong dipolar interactions and high densities also quickly take trapped molecular samples into the hydrodynamic regime, causing collective behavior to crossover from gaseous kinetic to fluid-like \cite{Kavoulakis98_PRA, Wang22_PRA0, Wang22_PRA, Wang23_PRA0, Wang23_PRA}.
Further, the description in this chapter has emphasized only non-resonant scattering in effective potentials as shown.  It is also possible that these potentials possess field-linked resonant states near the scattering threshold \cite{Avdeenkov03_PRL, Avdeenkov04_PRA, Chen23_Nat, Chen24_Nat, Li25_arxiv}, which would strongly influence the cross sections and rethermalization dynamics.  This area has yet to be explored.
The rapid experimental progress in controlling ultracold polar molecules has quickly established them as a powerful resource for quantum science, showing promise for quantum simulation \cite{Gorshkov11_PRL, Gorshkov11_PRA, Yan13_Nat, Christakis23_Nat, Carroll25_Sci, Miller24_Nat}, quantum information processing \cite{Zhang22_PRXQ, Wang22_PRXQ, Holland23_Sci, Guttridge23_PRL, Ruttley24_PRXQ, Ruttley25_Nat}, and possibly exotic quantum applications that have yet to be envisaged.

\section{ Acknowledgements }

Many of the insights on ultracold molecules presented here have come from invaluable discussions with T. Karman, I. Stevenson, S. Will and J. Ye. 
The authors thank S. I. Mistakidis for a thorough reading of this chapter. 
JLB Acknowledges support from the JILA Physics Frontier Center, grant NSF-2317149. 
RRWW acknowledges support at ITAMP by the National Science Foundation. 
This chapter will appear in the Springer book entitled ``Short and Long Range Quantum Atomic Platforms - Theoretical and Experimental Developments".

% \clearpage
% \bibliographystyle{unsrtnat}	% or "siam", or "alpha", etc.
% \nocite{*}		% list all refs in database, cited or not
\bibliography{main.bib}		% Bib database in "refs.bib"

\end{document}